\documentclass[twocolumn,floatfix,aps,prc,showpacs,preprintnumbers,amsmath,amssymb]{revtex4}
\usepackage{epsfig}
\usepackage{dcolumn}

\begin{document}

\title{ Precise measurement of the $^7$Be(p,$\gamma$)$^8$B S-factor}

\author{A. R. Junghans}
\altaffiliation{present address: Forschungszentrum Rossendorf, Postfach 510119, 01324 Dresden, Germany}
\altaffiliation{electronic address: A.Junghans@fz-rossendorf.de}
\author{E. C. Mohrmann}
\altaffiliation{electronic address: mohrmann@u.washington.edu}
\author{  K. A. Snover}
\altaffiliation{electronic address: snover@gluon.npl.washington.edu}
\author{T. D. Steiger}
\altaffiliation{present address: Cymer, Inc., 17075 Thornmint Ct., San Diego, CA 92127.}
\author{E. G. Adelberger}
\author{J. M. Casandjian}
\altaffiliation{permanent address: GANIL, B.P. 5027, 14021 Caen Cedex, France.}
\author{H. E. Swanson}
\affiliation{Center for Experimental Nuclear Physics and Astrophysics, University of  Washington,
Seattle,~Washington~98195}

\author{L. Buchmann}
\author{S. H. Park}
\altaffiliation{Dept. Physics, Seoul Nat. Univ., Seoul 151-742, Rep. of Korea.}
\author{A. Zyuzin}
\author{A. Laird}
\affiliation{TRIUMF, 4004 Wesbrook Mall, Vancouver, B.C., Canada V6T 2A3}

\date{\today}

\begin{abstract}

We present new measurements of the $^7$Be(p,$\gamma$)$^8$B cross section from $\bar{E}_{\rm cm}$ = 116 to 2460 keV, that incorporate several improvements over our previously published experiment, also discussed here.  Our new measurements lead to  S$_{17}$(0) = 22.1 $\pm$ 0.6(expt) $\pm$ 0.6(theor)  eV b based on data from $\bar{E}_{\rm cm}$ = 116 to 362 keV, where the central value is based on the theory of Descouvemont and Baye.  The theoretical error estimate is based on the fit of 12 different theories to our low energy data.  We compare our results to other S$_{17}$(0) values extracted from both direct ($^7$Be(p,$\gamma$)$^8$B) and indirect (Coulomb dissociation and heavy-ion reaction) measurements, and show that the results of these 3
types of experiments are not mutually compatible.  We recommend a ``best" value, S$_{17}$(0) = 21.4 $\pm$ 0.5(expt) $\pm$ 0.6(theor) eV b, based on the mean of all modern direct measurements below the 1$^+$ resonance.  We also present S-factors at 20 keV which is near the center of the Gamow window: the result of our measurements is S$_{17}$(20) = 21.4 $\pm$ 0.6(expt) $\pm$ 0.6(theor)  eV b, and the recommended value is S$_{17}$(20) = 20.7 $\pm$ 0.5(expt) $\pm$ 0.6(theor) eV b.
\end{abstract}

\maketitle

\section{Introduction}

Studies of high-energy neutrinos produced by $^8$B decay in the sun have shown that when these neutrinos
are detected on earth the number that remain electron neutrinos ($\nu_e$'s) is only about half of the number that have oscillated into
other active species~\cite{sno}.
Precise predictions of the {\em production rate} of $^8$B solar neutrinos are important for testing solar models, and for limiting the allowed neutrino mixing parameters including
possible contributions of sterile species.

The predicted $^8$B production rate~\cite{bahcall} is based on solar-model calculations
that incorporate measured reaction rates for each of the solar burning steps following the initial $p+p$ reaction,
the most uncertain of which is the $^7$Be(p,$\gamma$)$^8$B rate. The currently recommended value for S$_{17}$(0), the astrophysical S-factor for this reaction, is $19^{+ 4}\!\!\!\!\!\!_{- 2}$ eV b~\cite{adelberger}.  However,  S$_{17}$(0) must be known to better than $\pm 5\%$ in order that its uncertainty not be important in the predicted $\nu_e$ production rate~\cite{bahcall}.

Because of the key role of S$_{17}$(0), it has been measured many times using a variety of techniques:
direct studies of the $^7$Be(p,$\gamma$)$^8$B cross section~\cite{kavanaghold,parker,kavanagh,vaughn,wiezorek,filippone,hammache,gialanella,hass, strieder,junghans, baby}, and indirect studies using Coulomb-breakup~\cite{motobayashi,kikuchi,iwasa,davids,schumann} or peripheral
heavy-ion transfer and breakup~\cite{azhari,trache} reactions.
The direct technique has the advantage that it studies the reaction that actually occurs in the sun, but is difficult experimentally because it requires a radioactive target or beam, and the cross sections are small.

Our direct measurements reported here incorporated several improvements over traditional methods.  We eliminated a major uncertainty in
many previous experiments due to uncertain and nonuniform target areal density by using
a $\sim$1 mm-diameter beam magnetically rastered to produce a nearly uniform flux
over a small $\sim$3.5 mm diameter target. We directly measured the energy-loss profile
of the target using a narrow $^7$Be($\alpha,\gamma)^{11}$C resonance.  We made frequent $\it{in}$ $\it{situ}$ measurements of the $^7$Be target activity to determine the target sputtering losses.  We also made the first (and, to date, the only) measurements of $^8$B backscattering losses.

This paper reports the results of three separate experiments that determine the $^7$Be(p,$\gamma$)$^8$B cross section over a range of 3 orders of magnitude. The first~\cite{junghans} used a target of 106 mCi initial activity (here called BE1~\cite{be22label})
to measure the $^7$Be(p,$\gamma$)$^8$B cross section at mean center-of-mass proton energies, $\bar{E}_{\rm cm}$ = 186 to 1203 keV, including the M1 resonance near $\bar{E}_{\rm cm}$ = 630 keV.  The $\alpha$-detector solid angle in this experiment was determined using a $^7$Li(d,p)$^8$Li reaction yield ratio.  In a second, abbreviated measurement with a new $^7$Be target of similar activity (called BE2), we determined the cross section in the range $\bar{E}_{\rm cm}$ = 876 to 2459 keV.  In this second measurement, we became concerned over possible inaccuracies in the $^7$Li(d,p)$^8$Li reaction yield ratio method for determining the $\alpha$-detector solid angle, which relied on calculated extrapolations of the continuous $\alpha$-spectrum tail below the experimental cutoff (a similar but smaller extrapolation is necessary to interpret the $^7$Be(p,$\gamma$)$^8$B reaction yields).

Recently we completed a $^7$Be(p,$\gamma$)$^8$B measurement with a 340 mCi target (called BE3) that incorporated several improvements over the BE1 experiment, including a thinner Si surface-barrier detector for which the correction (and hence the uncertainty) due to the fraction of the $^8$B $\beta$-delayed $\alpha$-spectrum lost below the detector threshold was minimal. We avoided using the $^7$Li(d,p)$^8$Li reaction for solid angle normalization, using instead a custom-made $^{148}$Gd $\alpha$-source fabricated on a backing of the same design as the one used for the $^7$Be target.   As a result, we were able to measure and minimize all important sources of systematic error in determining the $^7$Be(p,$\gamma$)$^8$B cross section.
This BE3 measurement, which covered the range $\bar{E}_{\rm cm}$ = 116 to 1754 keV, yields our best determination of the absolute cross section.  Hence we base our absolute S-factor determination on this experiment.  Our new determination of S$_{17}(0)$ turns out to be in excellent agreement with our previously published value~\cite{junghans}.

We discuss all 3 experiments in Secs.~\ref{experimentalprocedure} - \ref{be1results} below, with an emphasis on the BE3 measurements.  Comments on the BE1 and BE2 experiments are labeled explicitly; unlabeled comments refer to the BE3 experiment.  We estimate the extrapolation uncertainty in S$_{17}(0)$ in Sec.~\ref{extrapolation}, and make detailed comparisons with other direct experiments in Sec.~\ref{directcompare}. We discuss indirect measurements in Sec.~\ref{indirectcompare} and in Sec.~\ref{recommendvalue} we recommend a ``best" value for S$_{17}(0)$.  We summarize our results in Sec.~\ref{summary} and we discuss their implications for solar and neutrino physics in Sec.~\ref{neutrinoimplications}.

\section{Experimental procedure}
\label{experimentalprocedure}
\subsection{Target fabrication}
Our $^7$Be targets were made at TRIUMF using a technique described previously~\cite{zyuzin1,zyuzin2}.  Briefly, the $^7$Be was produced by the $^7$Li(p,n)$^7$Be reaction at E$_p$ = 13 MeV, using a Li metal target and the TR13 cyclotron at TRIUMF.  The $^7$Be was chemically separated using a glass filter and various solutions including HCl, then heated and dried, leaving the $^7$Be predominantly in the form of $^7$BeO.  Then a 2-step reduction/distillation was performed in vacuum.  In the first step, the $^7$BeO was heated in a Zr-lined Mo crucible, depositing $^7$Be metal on a  Mo piece which served as the lid of the apparatus.  In a second distillation the Mo piece served as the heating crucible and the $^7$Be was deposited on a target backing.   This technique yielded  $^7$Be targets of high-purity (40 - 63\% by atom number - see Secs.~\ref{alphagamma} and \ref{backscattering}), with initial activities of 106 mCi (BE1), 112 mCi (BE2) and 340 mCi (BE3).  The $^7$Be activities produced at TRIUMF for these targets were roughly 220, 420 and 630 mCi, respectively.

The target backings consisted of 1.3 by 1.5 cm Mo plates with stainless-steel water-cooling tubes brazed onto their back sides, and mounting brackets on the top for attaching the backing to the rotating arm (see Fig. 4 of \cite{zyuzin2}).  The front face of the plate consisted of a flat surface with a 4 mm diameter, 1.5 mm high Mo post in the center.  Prior to $^7$Be deposition, a Mo washer was press-fitted around the post.  The washer and post were machined to precise tolerances to ensure a tight fit, and the post-plus-washer was machined flat after assembly.  After $^7$Be deposition, the washer was broken away, leaving $^7$Be only on the post.  This procedure ensured that $^7$Be remained only within a small $\approx$3.5 mm diameter central area, a feature that was important for our large-area-beam/small-area-target technique described below.

\begin{figure}
\includegraphics[angle=270,width=0.5\textwidth]{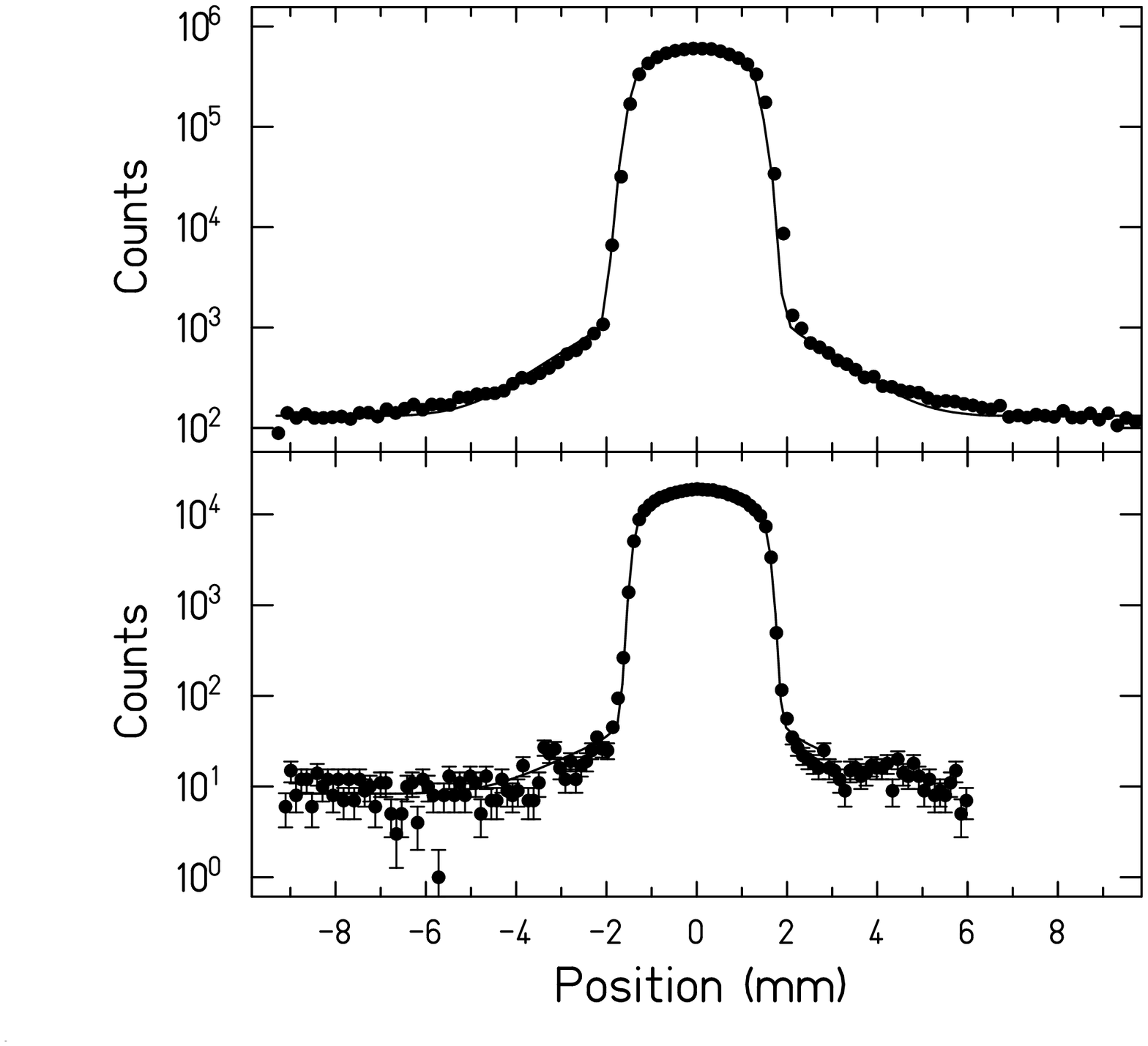}
\caption{BE3 Activity scan measured before (top panel) and after (bottom panel) the $(p,\gamma)$ experiment. Solid curves - calculated activity (see text).}
\label{activity_scan}
\end{figure}

A $\gamma$-activity scan of the BE3 target is shown in Fig.~\ref{activity_scan}.  The scan was measured with a Ge detector collimated by a 51 mm thick ``heavy metal" (tungsten alloy) block  containing a 0.125 mm slit (see also ref.~\cite{zyuzin2}).  Scans measured both before and after the $^7$Be(p,$\gamma$)$^8$B experiment showed very similar shapes.  Scans measured with the target turned sideways showed that an insignificant  amount of $^7$Be was located elsewhere than on the top of the post, and were used to determine the position-resolution function for a line source (see ref.~\cite{zyuzin2}).  Fits to the measured scans using this resolution function showed that the $^7$Be density distribution on the post was constant within a radius of 1.5 $\pm$ 0.1 mm and decreased  to zero at a radius of 1.8 $\pm$ 0.1 mm.

Additional information on target purity and composition is given in Secs. \ \ref{alphagamma} and \ \ref{backscattering} below.

\subsection{Experimental apparatus}
The $^7$Be(p,$\gamma$)$^8$B cross sections were measured at the University of Washington Center for Experimental Nuclear Physics and Astrophysics, using p, d and $\alpha$ beams at energies up to 3 MeV from the tandem Van de Graaff accelerator with a terminal ion source.   For the BE1 and BE2 experiments, proton beam currents up to 16 $\mu$A  were used.  For the BE3 experiment a straight-field accelerating tube was installed in place of the first high-energy spiral-field tube section to obtain lower energies and proton beam currents up to 35 $\mu$A.   Typical $\alpha$ beam currents were 8 $\mu$A or less on target. Based on thermal tests of our water-cooled backings, beam-power levels on target were kept to less than 10 watts
to prevent target damage from beam heating.

\begin{figure}
\includegraphics[width=0.5\textwidth]{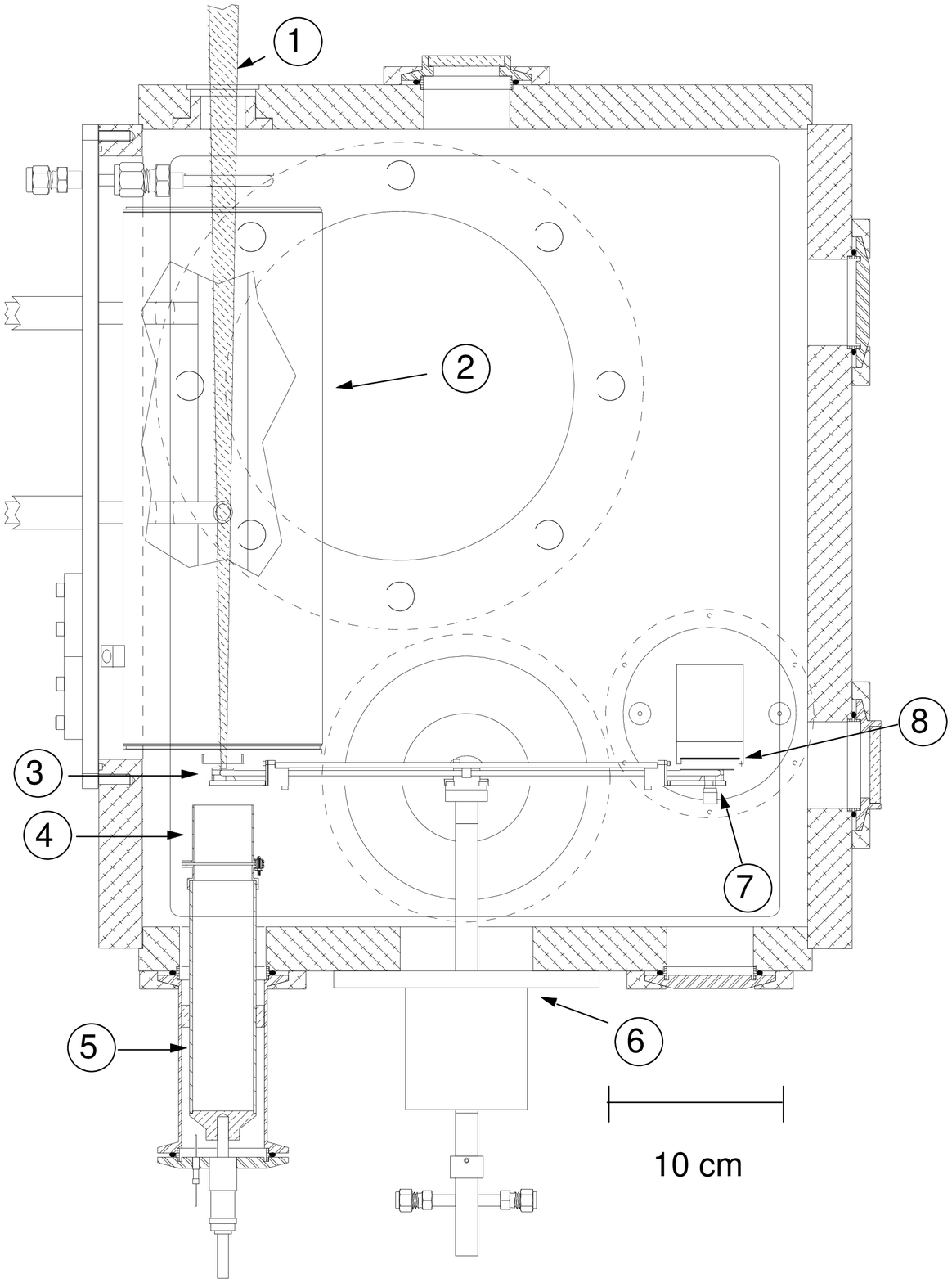}
\caption{Top view of target chamber.  1: shaded beam profile;  2: LN$_2$ cold trap; 3: target; 4: electron suppressor; 5: Faraday cup; 6: ferrofluidic seal; 7: aperture plate; 8: $\alpha$-detector.  The 3 flanges in the bottom plate (from large to small size) were for pumping, heavy-metal shield and the $\alpha$-detector mount.}
\label{chamber}
\end{figure}

Figure~\ref{chamber} shows a top view of the target chamber. The beam,  indicated by the shaded taper, enters the chamber from the upper left, and passes through a large-area
aperture followed by a 31 cm-long cylindrical LN$_2$ cold trap with inner and outer diameters of 2.9 and 11.4 cm.  The cold trap had a removable Cu liner which captured most of the $^7$Be sputtered from the target during long bombardments.  In Fig.~\ref{chamber} the beam is shown striking the target mounted on one end of a water-cooled rotating arm. A Cu sleeve extended from the cold trap to within $\sim$1 mm of the target to shield it from condensible vapors.  The arm was mounted on a shaft connected via a ferrofluidic seal to a computer-controlled servo motor.  This shaft also carried the water flow for the arm cooling.  A 4 cm by 10 cm copper plate was mounted on the opposite end of the arm from the target.  This plate contained 4 apertures with nominal diameters of 1, 2, 3 and 4 mm, spaced 1.8 cm between centers.  The apertures were sized by pressing precision-machined steel balls through slightly undersized, machined holes, and the aperture areas were measured to a precision of  $\pm$ 0.2\% or better using a high-magnification traveling microscope.  When the target was in the bombardment position, the aperture plate was directly in front of a Si surface-barrier detector as illustrated.
Rotating the arm by +180$^{\circ}$ from its target bombardment position placed the 3 mm aperture in the beam and put the target in front of the Si $\alpha$-detector, where the $^8$B $\beta$-delayed $\alpha$'s were counted.  In this configuration, the beam that passed through the aperture was collected in a Faraday cup, shown in Fig.~\ref{chamber} together with its electron suppressor ring.  A subsequent
$-180^{\circ}$ rotation placed the target back in the bombardment position.

A $^{148}$Gd $\alpha$-source with a moveable shutter was located near the Si detector and used to monitor the detector gain when the arm was oriented vertically.
Several thin Al fixed shields (not shown in Fig.~\ref{chamber}) were placed in the chamber and a curved shield was mounted on the aperture end of the arm to prevent scattered proton beam from reaching the $\alpha$-detector.  With this shielding in place, no scattered beam was detected above the thresholds employed in the $\alpha$-spectrum measurements.

In order to suppress carbon buildup on the target, the chamber was evacuated using a system designed to suppress volatiles.  A helium cryopump was used for high-vacuum pumping, and sorption pumps for roughing. The vacuum system was designed in part for radiation safety.  $^7$Be targets were transferred in and out of the chamber using a portable Pb-shielded transfer device that allowed the target to be mounted on the arm with remote handling rods.  Once the target was mounted, the arm could be rotated vertically downward and a heavy-metal shield directly below the arm could be raised around the target to suppress the 478 keV $\gamma$-radiation when the target was not in use.  A collimated Ge detector was mounted on the chamber lid and used to make frequent $\it{in}$ $\it{situ}$ measurements of the $^7$Be target activity.
A large, shielded NaI spectrometer, located on one side of the
chamber with its front Pb collimator $\sim$ 30 cm from the target, was used for $^7$Be($\alpha, \gamma)^{11}$C measurements of the target energy loss profiles and $^{19}$F(p, $\alpha \gamma)^{16}$O resonance measurements of the accelerator energy calibration.
Horizontal and vertical magnetic deflection coils for beam rastering were located 1.1 m upstream of the target, and a magnetic quadrupole lens was located 2 m upstream.

\subsection{Uniform beam-flux technique}

In an in-beam experiment, the yield $Y$ is related to the cross section $\sigma$, the target areal density $dN/dA$ and the beam current density $dI/dA$ by the expression
\begin{equation}
Y = \frac{\sigma}{q} \int \frac{dN }{ dA} \frac{dI }{ dA} dA~,
\label{int}
\end{equation}
where the integral is over the target area and $q$ is the charge of a beam particle.
If the target areal density is constant, and the area of the target is larger than the area of the beam, this equation reduces to the usual expression
\begin{equation}
Y = \frac{\sigma}{q} \frac {dN}{ dA} I,
\label{int1}
\end{equation}
where $I$ is the total beam current.  Eq. \ (\ref{int1}) is commonly used as an approximation in cases where the areal density of the target is not constant.  However, in this case it is difficult to know the error involved, since it depends on both the target areal density $dN/dA$ and beam density $dI/dA$ nonuniformities, which are often not known.

On the other hand, if the beam current density is constant and the target area is smaller than the beam area, Eq.\ (\ref{int}) reduces to
\begin{equation}
Y = \frac{\sigma}{q}  N \frac{dI}{ dA},
\label{int2}
\end{equation}
where $N$ is the total number of target atoms.  This method, which we used, offers several advantages over the conventional procedure.
We determined $N$ precisely by measuring the $^7$Be activity, and $dI/dA$ by measuring the beam transmission through different sized apertures~\cite{workshop}.   Nevertheless, in any practical experiment the beam density can never be made completely uniform.  To understand
possible errors in using Eq.\ (\ref{int2}) we needed to know the density nonuniformities of both the beam and target, or more precisely the nonuniformity in the product $(dN/dA)(dI/dA)$.  We achieved this by relative $^7$Li(d,p)$^8$Li measurements as a function of raster amplitude, as described in Sec.\  \ref{beamtargetuniformity} below.

We made use of the $t_{1/2}$ = 770 $\pm$ 3 ms half-life of $^8$B by bombarding the target and then rotating it away from the beam to a shielded location where the $^8$B decays were detected.  A single measurement corresponded to a
large number of such cycles.  Modifying Eq.\ (\ref{int2}) to include this process, and making all relevant factors explicit, we obtain
\begin{equation}
\label{cross}
\sigma(\bar{E}_{\rm cm}) = \frac{Y_{\alpha}(E_p) F_{\alpha}(E_p) \beta(^8\mbox{B})}
{2\phi_p N_{\text{Be}}(t)
\Omega /4\pi}
\end{equation}
where $\bar{E}_{\rm cm}$ is discussed below, $E_p$ is the bombarding energy,
$Y_{\alpha}(E_p)$ is the $\alpha$ yield above a threshold energy,
$F_{\alpha}(E_p)$ is a correction for the fraction of the $\alpha$-spectrum
that lies below the threshold,  $\phi_p$ is the integrated number of beam protons per
cm$^2$, $N_{\text{Be}}(t)$ is the number of $^7$Be atoms present at time $t$ of the measurement,
and $\Omega$ is the solid angle of the $\alpha$-detector.

The factor $\beta (^8$B) is the inverse of the timing efficiency for counting $^8$B decay and is given by
\begin{equation}
\beta (^8\mbox{B}) = \frac{\lambda t_1 [1 - e^{-\lambda (t_1 + t_2 + t_3 + t_4)}] }{ (1 - e^{\lambda t_1})[e^{-\lambda t_2} - e^{-\lambda (t_2 + t_3)}]},
\label{beta}
\end{equation}
where $\lambda = 0.693/t_{1/2}(^8$B), $t_1 = 1.50021 \pm 0.00023$ s is the bombardment period, $t_2 = 0.24003 \pm 0.00004$ s is the transfer time to the counting position, $t_3$ = $t_1$ is the counting period and $t_4 = 0.26004 \pm 0.00004$ s is the transfer time to the bombardment position.  The arm rotation was controlled by a hardwired electronic sequencer box, and the time periods were measured with a precision pulser.   Additional tests
assured us that the actual arm movement and beam on/off periods were consistent with these times, with uncertainties similar to or less than those listed above, so that $\beta (^8$B) = 2.923 $\pm$ 0.006.

\subsection{Beam rastering and current integration}

A highly uniform beam flux on target was achieved by rastering the beam in x and y using the magnetic deflection coils.  First, the beam was tuned through the 1 mm aperture, with typical transmission 60\% or better.  Then the beam was uniformly rastered over an approximately square area, typically about 7 mm x 7 mm, by driving the rastering coils with a triangular voltage waveform.  Incommensurate rastering drive frequencies of 19.03 Hz and 43.00 Hz were selected to minimize beating irregularities in the beam transmitted through a small aperture.  Pickup
coils monitored the time dependence of the magnetic rastering fields.  The rastering drive voltages used for each cross section measurement were determined from aperture scans as described below.

The arm was electrically isolated from the chamber, and biased to +300 V to minimize secondary electron losses when the beam struck either the target or the aperture plate.  The Faraday cup electron suppressor was biased to $-300$ V.  The beam currents striking both the Faraday cup and the arm were integrated separately, and were recorded during both the bombardment and counting phases.  Our notation is: $Q_{T}$ - integrated beam striking the target during the bombardment phase;  $Q_{A}$ - integrated beam striking the aperture during the counting phase; and $Q_{C}$ - integrated beam passing through the aperture and striking the cup during the counting phase.  Just before (after) arm rotation, the beam was rapidly swept away from (back onto) the target using one of the magnetic deflection coils.

Our primary integrated beam-charge reference was $Q_{C}$, accumulated during the counting phases of a run (many complete arm-rotation cycles).  Thus $(1/q)Q_{C}/A$, where $A$  is the area of the 3 mm aperture, is a measure of the flux factor $\phi_p$ in Eq.\ (\ref{cross}) (see below).   There are two assumptions here: first, that the integrated beam flux striking the slightly larger diameter target was equivalent to $Q_{C}/A$.  Secondly, that the integrated beam flux striking the target during the bombarding phases of a run was equivalent to $Q_{C}/A$ - that is, as a result of time-averaging, the integrated beam flux passing through the 3 mm aperture and striking the cup during the counting period was the same as the integrated beam flux striking the
target during the bombardment period.  The first assumption, related to beam (and target) uniformity, was tested as described in Sec.~\ref{beamtargetuniformity} below.

We tested the second assumption by computing the factor $Q_{T}\cdot Q_{C}/(Q_{C} + Q_{A}$).  Here $Q_{C}/(Q_{C} + Q_{A})$ is the fraction of the beam that passes through the 3 mm aperture, $Q_{T}$ is the total beam striking the target, and thus $Q_{T}\cdot Q_{C}/(Q_{C} + Q_{A})$  measures the (integrated) fraction of the beam striking the central 3 mm region of the target.

In a series of test runs, we found the ratio of the ``good geometry" (Faraday cup) and ``poor geometry" (biased arm) beam-flux factors $Q_{C}$ and $Q_{T}\cdot Q_{C}/(Q_{C} + Q_{A}$) differed from unity by  0.02\% $\pm$ 0.8\%, where the uncertainty was determined from the run-to-run fluctuations.  We also found that for short, high-yield $^7$Li(d,p)$^8$Li diagnostic runs, the reaction yields showed smaller run-to-run fluctuations when normalized to $Q_{T}\cdot Q_{C}/(Q_{C} + Q_{A})$ than when normalized to $Q_{C}$.  Hence we adopted the
normalization $\phi_p$ = $(1/q)Q_{T}\cdot (Q_{C}/A)/(Q_{C} + Q_{A}$) based on the agreement between the good-geometry and poor geometry results.
We made two additional checks on the accuracy of our beam-current integration.  We varied the cup suppressor bias in the range -300V $\pm$ 45V, and found $<$0.5\% change in beam current.  We also set a limit of 10$^{-4}$ on the neutral H content of the beam by first tuning a proton beam onto a LiF target and measuring the $^{19}$F(p,$\alpha \gamma$)$^{16}$O reaction yield at the 340 and 484 keV resonances, then sweeping the H$^+$ away from the target
with one of the magnetic deflectors and repeating the measurement.
In this manner we assigned an overall systematic uncertainty of $\pm$ 0.9\% on the integrated beam flux.

\subsection{Beam and target uniformities}
\label{beamtargetuniformity}

\begin{figure}
\includegraphics[width=0.5\textwidth]{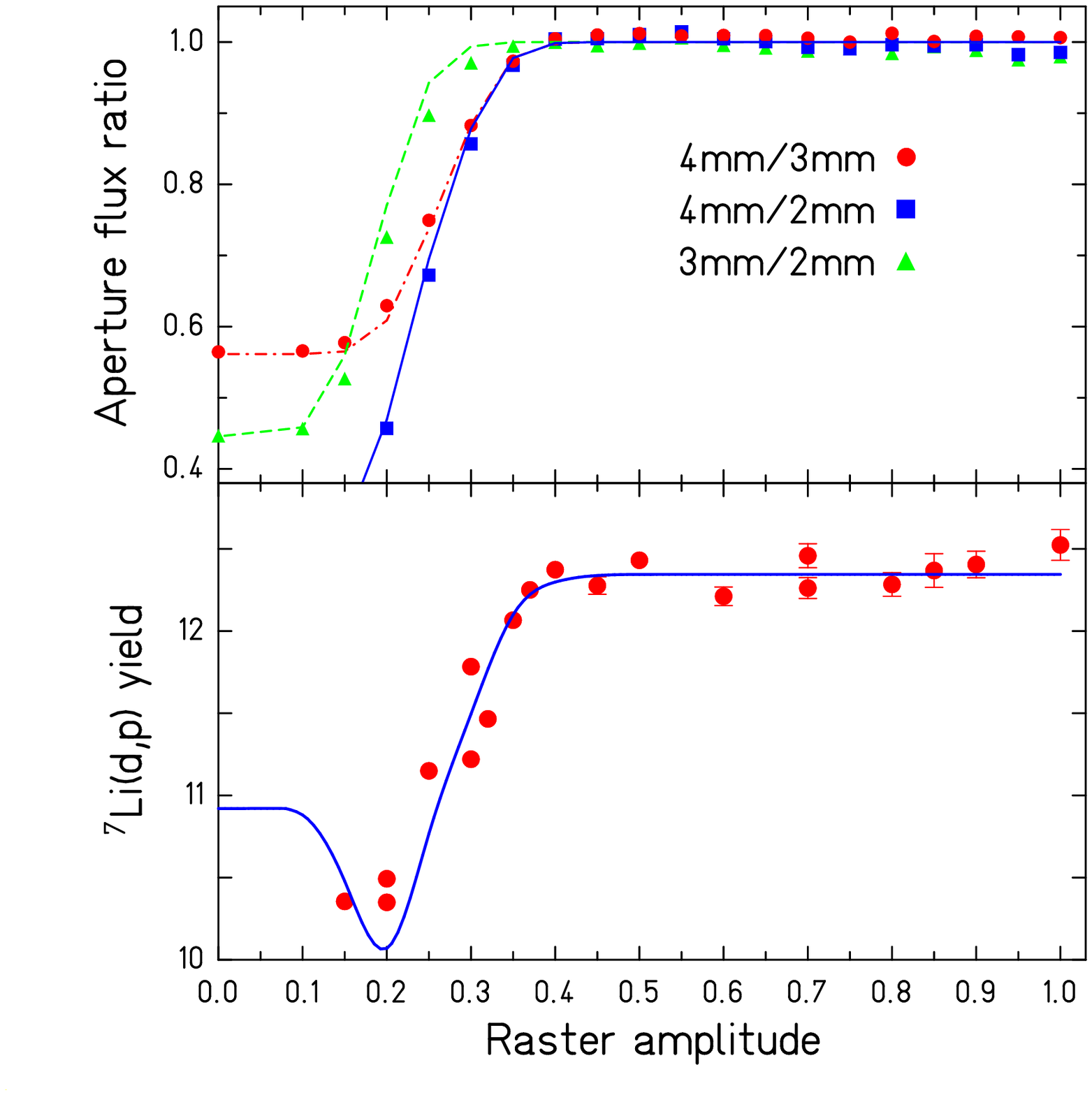}
\caption{Top panel: 770 keV deuteron beam transmission
ratios through different apertures, vs. raster amplitude (relative units).  Bottom panel:
$^7$Li(d,p)$^8$Li yield at 770 keV, normalized to the integrated beam flux through the 3 mm
aperture, vs. raster amplitude, measured with the same tune as the aperture ratio data.  The curves are described in the text.}
\label{dp}
\end{figure}

Figure~\ref{dp} shows measurements made during the BE1 experiment to determine the beam and beam-target uniformity.
The beam uniformity was determined by measuring the transmissions through
2, 3 and 4 mm apertures as functions of (approximately equal) amplitudes of the x and y
triangular raster waveforms.  The top panel of Fig.~\ref{dp} shows measurements made  with a 770 keV deuteron beam, and curves calculated by folding a Gaussian beam spot
with a rectangular raster distribution.
The uniformity of the product of the beam and target densities was determined by
the raster-amplitude dependence of the $^7$Li(d,p)$^8$Li yield from the $^7$Be
target at $E_d$= 770 keV, shown in the bottom panel of Fig.~\ref{dp}.
The curve is a convolution of the target density estimated from $\gamma$-activity scans, and beam
profile determined by the transmission ratios, including a fitted aperture-target misalignment of 0.5 mm.
The behavior at small raster amplitude was due to this misalignment.  The point at which this yield flattened out determined
the minimum safe raster amplitude, and is similar to the point at which the
aperture ratio data flattened out.  We chose 0.42 as the safe raster amplitude for 770 keV
deuterons, and assigned a conservative $\pm 1\%$ nonuniformity uncertainty here.
Aperture transmission curves were measured at most proton energies and used, with reference to Fig.~\ref{dp}, to determine the
raster amplitude for each energy and tune so that the beam-target
nonuniformity was $< 1\%$.  Independent estimates of the safe raster amplitudes
were made by folding the target-density distribution~\cite{zyuzin2}
with beam-flux distributions determined from the proton aperture-transmission data, with results consistent with the above procedure.  The same procedure was followed in the BE2 and BE3 experiments.

\subsection{$^7$Be activity measurements}

\begin{figure}
\includegraphics[width=0.5\textwidth]{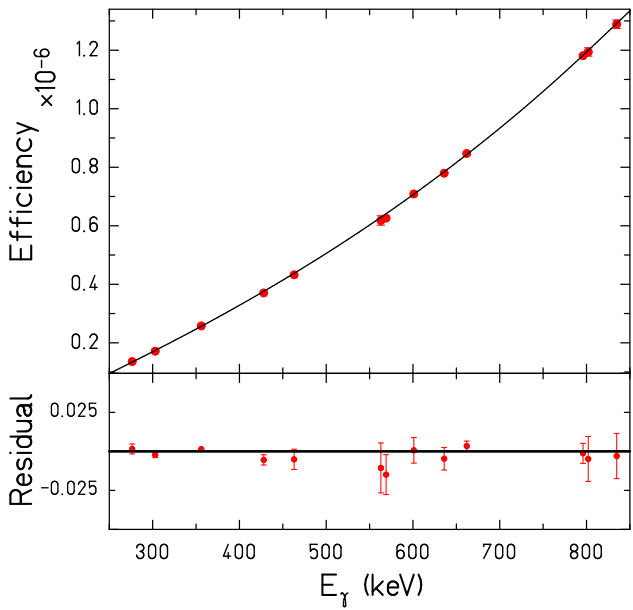}
\caption{Top panel:  $\it{in}$ $\it{situ}$ Ge detector efficiency calibration.  The curve is a third order polynomial fit. The increase of efficiency with energy is due to the absorber.  Bottom panel: residuals of the fit. }
\label{Gecalib1}
\end{figure}

The $^7$Be activity was measured $\it{in}$ $\it{situ}$ in a ``close" geometry (target arm vertical and a source-to-detector distance of $\approx$ 27 cm) by counting the 478-keV $\gamma$-rays with a 50\% efficient Ge detector  mounted on the lid of the target chamber.  An absorber consisting of 8 cm of Al plus 5 cm of steel
reduced the counting rate in the Ge detector to 1.5 kHz or less.  A cylindrical Pb collimator 6 cm long, with a 2 cm hole in the center ensured that the detector could not view the small amount of $^7$Be sputtered from the target and deposited mainly on the cold trap.  We assumed the accepted $^7$Be decay values
of $t_{1/2}$ = 53.12 $\pm$ 0.07 d, and $BR$ = 10.52 $\pm$ 0.06\% for the decay branch to the 478 keV level~\cite{halflives}.
The Ge detector efficiency, $\epsilon_{478}$, was determined using radioactive sources mounted on the arm in the same position as the $^7$Be target.  We fit
13 lines in the range 276 to 835 keV from $^{125}$Sb, $^{134}$Cs, $^{133}$Ba, $^{137}$Cs and $^{54}$Mn
sources calibrated typically to $\pm 0.8\% (1\sigma)$~\cite{isotope},
and obtained  $\chi^2 / \nu$ = 1.2.   Figure~\ref{Gecalib1} shows the calibration.

We made an independent check of the Ge detector calibration using a second $^{137}$Cs source calibrated
independently to $\pm 0.4\% (1\sigma)$~\cite{french}.  The result confirmed the correctness of our $\epsilon_{478}$ determination at the level of (0.4 $\pm$ 0.8)\% .

In addition, we made a separate determination of the BE3 target activity approximately 2 months after the end of the BE3 cross section measurements, using the Ge detector with a Pb collimator with a 1.6 cm diameter aperture, no absorbers, a source distance of about 200 cm and the same calibration sources used in the earlier measurements.  BE3 target activities inferred using this calibration and the 53.12 d half-life were 1.5 $\pm$ 1.5\% higher than those determined with the $\it{in}$ $\it{situ}$ calibration.  Since these results agree within errors,  we used the average of these 2 calibrations.

The number of $^7$Be atoms in Eqn.\ \ref{cross} is given by $N_{Be}(t)$ = (3.7 x $10^{10}$) $\cal A$ $t_{1/2}$/ln 2, where the target activity $\cal A$ = $N_{\gamma}/(3.7$ x $10^{10} \epsilon_{478} BR$) is given in Curies and $N_{\gamma}$ is the number of photopeak 478 keV counts/s in the Ge detector.

There is some evidence that the $^7$Be half-life depends weakly on host material (see e.g.  \cite{norman,ray,liu}).  We have taken the difference between the half-life in Au ( $\sim$53.31 d) and LiF (53.12 d) as representative of the uncertainty due to host material (see Table 1 of \cite{norman}), which introduces an additional $\pm$0.4\% uncertainty in the conversion of the measured activities to $N_{Be}(t)$.  The total systematic uncertainty in $N_{Be}(t)$ is $\pm$ 1.1\%.

We made a similar, second determination of the BE1 target activity, in this case 20 months after the BE1 $\it{in}$ $\it{situ}$ measurements.  Assuming a 53.12 d half-life, these two BE1 activity determinations are in good agreement, differing by 0.3 $\pm$ 1.9\%.  Using a 53.31 d half-life the difference was 2.7\%, suggesting that the $^7$Be half-life in our host material is close to the accepted value.

\begin{figure}
\includegraphics[width=0.5\textwidth]{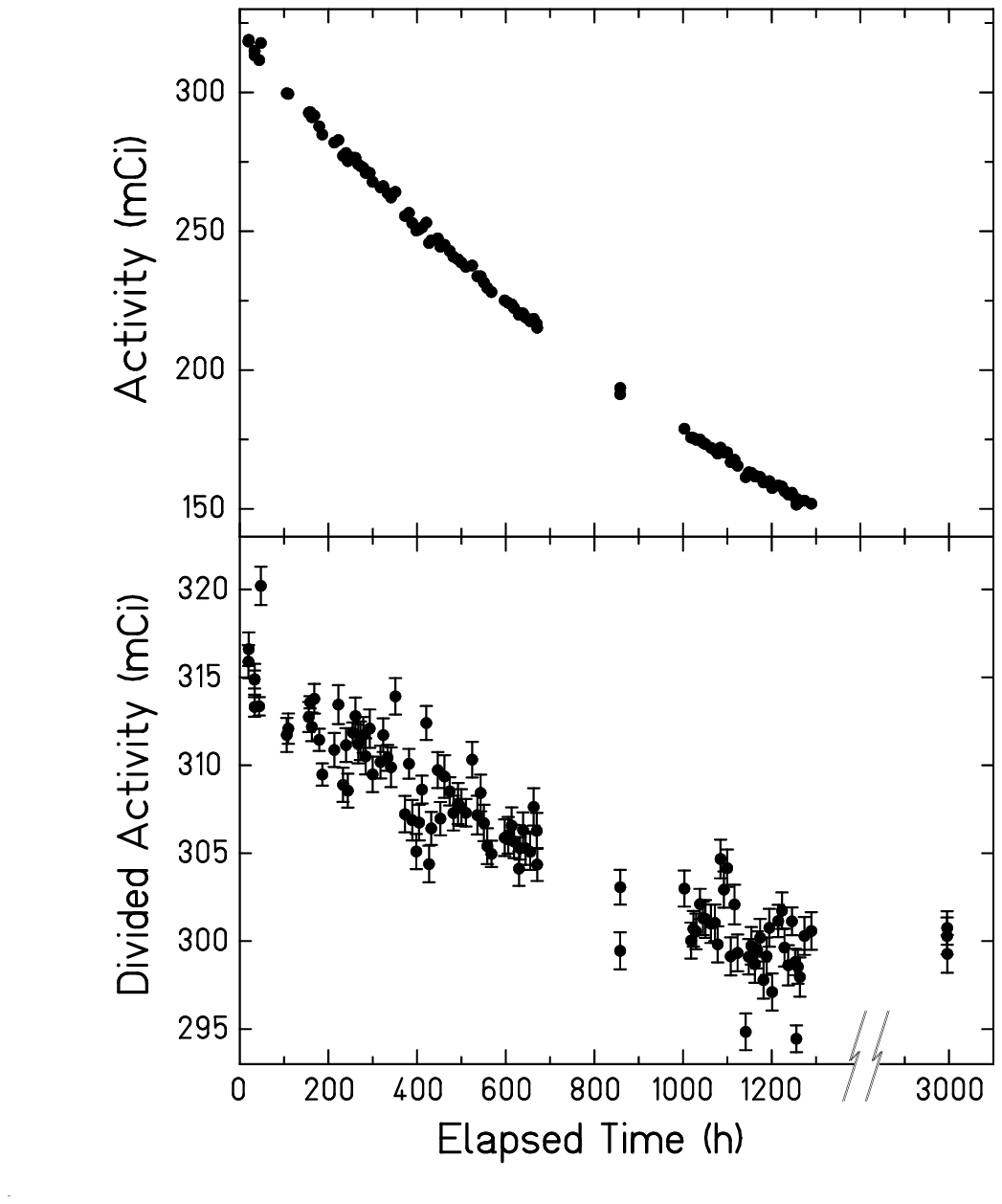}
\caption{Top panel: BE3 target activity versus time since the beginning of the $(p,\gamma)$ measurements.
Bottom panel: same as top panel but with the decay factor e$^{-\lambda t}$ divided out, where $\lambda$ = 0.693/53.12 d.
Note that the divided activity approximately 1700 hr after the end of the experiment agrees with the values found immediately after the end of the experiment. This demonstrates that the downward slope during the experiment was due to a loss of target material and not to an incorrect half-life.}
\label{activity}
\end{figure}

The $^7$Be activity measured during the BE3 experiment is shown in Fig.~\ref{activity}.  As can be seen from the lower panel, approximately 15 mCi was lost due to sputtering during the cross-section measurements.  Compared to a mean activity of  $\sim$230 mCi during the cross-section measurements, this is a 7\% loss.  Hence this activity monitoring was very important in both the relative and absolute cross section determinations.

\subsection{$^7$Be target energy-loss profiles}
\label{alphagamma}

In previous $^7$Be(p,$\gamma$)$^8$B experiments, the energy-loss widths of the targets were estimated from the broadening of the 41 keV wide $^7$Be(p,$\gamma$)$^8$B resonance at $E_p$ = 720 keV~\cite{hass,baby,strieder}, the 12 keV wide $E_p$ = 441 keV $^7$Li(p,$\gamma$)$^8$Be resonance~\cite{filippone}, or from a calculation using estimated amounts of contaminants~\cite{hammache}.

We determined the complete energy-loss profiles directly from measurements of the yield of the narrow ($\Gamma << $ 1 keV) $^7$Be($\alpha, \gamma)^{11}$C resonance~\cite{be7ag}, obtained using the large NaI spectrometer.  Since the resonance may be approximated as a $\delta$-function,  and the experimental energy resolution was very good, $\sim$ 1 keV, the measured $^7$Be($\alpha, \gamma)^{11}$C yield directly reflects the energy loss of the beam in the target, and may be converted into the corresponding energy loss distribution for incident protons of a given energy using ratios of known proton and $\alpha$-particle energy-loss functions.

The measured ($\alpha, \gamma)$ profiles were corrected for small backgrounds from cosmic rays and the $^9$Be($\alpha, n )^{12}$C reaction.  In the BE1 experiment, the beam-related background was estimated from the observed yield of 4.4 MeV $\gamma$-rays; in the BE3 experiment, by scaling measurements of the $\gamma$-ray spectrum from $^9$Be + $\alpha$ obtained with a $^9$Be target.

In the BE1 experiment, the resonance profile was measured in the middle and at the end of the (p,$\gamma$) measurements.  One of these profiles is shown in Fig. 2 of ref.~\cite{junghans}.
The two measured profiles agreed within experimental error, and the apparent resonance energy reproduced within 1 keV.  Thus these measurements show that the target profile did not change and there was negligible carbon buildup during the (p,$\gamma$) measurements.  We found the energy of this resonance to be E$_{\alpha}$ = 1378 $\pm$ 3 keV, in good agreement with the previously determined value of 1376 $\pm$ 3 keV~\cite{be7ag}.  The measured BE2 resonance profile was very similar to the E1 profiles discussed above.

\begin{figure}
\includegraphics[width=0.5\textwidth]{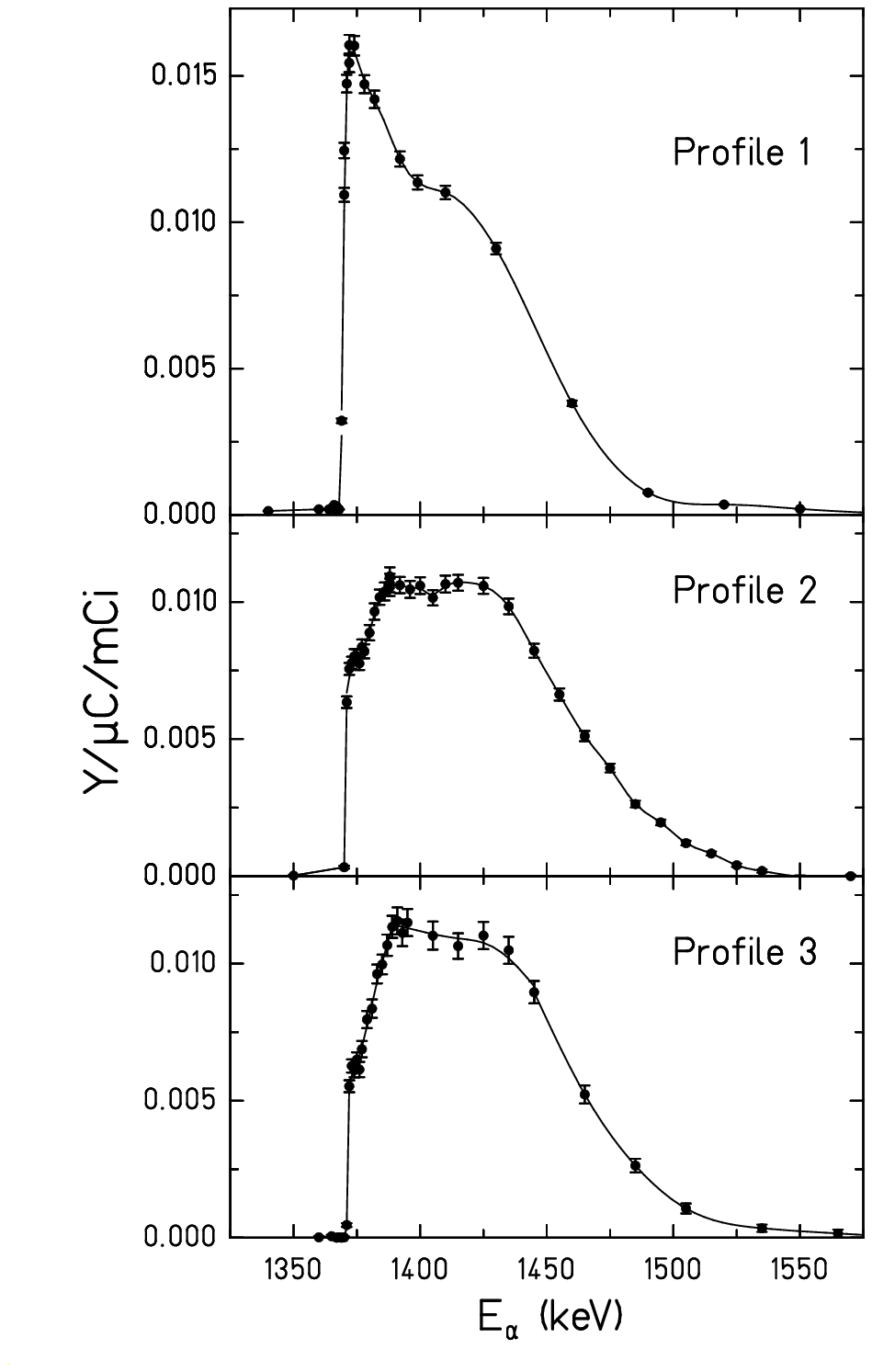}
\caption{$^7$Be($\alpha, \gamma)^{11}$C resonance profiles measured before (top panel), in the middle (center panel) and after (bottom panel) the $(p,\gamma)$ measurements with the BE3 target.  The graphs show the capture $\gamma$-ray yield per $\mu$C of $\alpha^+$ beam, per mCi of target activity, versus E$_{\alpha}$, and the curves are to guide the eye.  A small nonresonant (neutron) background from $^9$Be($\alpha, n \gamma)^{12}$C and a small beam-off background have been subtracted from each profile.}
\label{alphagammafig}
\end{figure}

BE3 target energy-loss profiles were measured three times: at the beginning, in the middle and at the end of the (p,$\gamma$) measurements, and are shown in Fig.~\ref{alphagammafig}.
Profiles 2 and 3 are very similar, and differ from profile 1.  An accident occured during the (p,$\gamma$) measurements at $E_p$ = 176.5 keV ($\bar{E}_{\rm cm}$ = 139.8 keV) in which the target was exposed to beam with a small raster amplitude.  The close similarity of the two profiles measured in the BE1 experiment, and profiles 2 and 3 in the BE3 experiment shows that under normal running conditions both of these targets were stable.  This provides a strong basis for associating the profile change with the raster accident.  The (p,$\gamma$) data taken during the accident were discarded, the data measured before the accident were associated with profile 1, and later data were associated with profiles 2 and 3.  Proton energy losses inferred from profiles 2 and 3 were the same within errors.  The leading edges of profiles 2 and 3 reproduced to better than $\Delta E_{\alpha}$ = 1 keV, and were located at 1 keV higher apparent alpha energy than the leading edge of profile 1.  This difference was within experimental error, indicating that C buildup was at most equivalent to $\Delta E_{\alpha}$ = 1 keV.

The observed profile widths, together with the measured $^7$Be activity, indicate the presence of target contaminants.  From the target-fabrication process~\cite{zyuzin1,zyuzin2} we expect contaminants with masses up to Mo.  Analysis of $^8$B backscattering measurements (see Sec.~\ref{backscattering}) shows that the contaminants are mostly Mo or a material with a similar Z.  From the profile widths and the known amount of $^7$Be present, we infer a $^7$Be:Mo stochiometry of 42:58 for BE1 and 63:37 for BE3 assuming no other contaminants  (see Sec.~\ref{backscattering}). These targets were much purer than, for example, those of ref.~\cite{filippone}.  Target properties are summarized in Table \ref{tabletargetprop}.
\begin{table}
\caption{Target properties.}
\label{tabletargetprop}
\begin{ruledtabular}
\begin{tabular}{lccc}
Target & BE1 & BE2  & BE3   \\
\hline
Initial activity (mCi) & 106 & 112 & 340 \\
Mean $\Delta E_{\alpha}$(keV) & 27 & 28 & 54\footnotemark \\
$^7$Be:Mo stochiometry & 42:58 & --\footnotemark & 63:37 \\
\end{tabular}
\end{ruledtabular}
\footnotetext[1]{Profile 2.}
\footnotetext[2]{No backscattering measurement.}
\end{table}

The measured ($\alpha, \gamma$) profiles were converted into proton energy-loss distributions $P(E_p)$ using the target composition discussed above, and ratios of energy-loss functions~\cite{srim} $dE_{\alpha}/dx$ and $dE_p/dx$,   ignoring straggling differences between protons and alphas.  Straggling, which is significant only in the region of the high-energy tail of the energy loss profile, was negligible in our case since the high energy tails of our measured profiles decreased about a factor of 10 more slowly than expected based on straggling calculations~\cite{bichsel}, due presumably to target nonuniformity.

Energy-averaged quantities were computed according to
\begin{equation}
\langle f(E_p) \rangle =   \frac{\int f(E_p)P(E_p)dE_p }{ \int P(E_p)dE_p}.
\label{energyaverage}
\end{equation}
where $f(E_p)$ is either $E_p$ or $\sigma(E_p)$.
Energy-averaging calculations are discussed further in Sec.~\ref{be3results} and \ref{be1results}.

\subsection{Alpha-detector solid angles}

In the BE3 experiment, $\alpha$-particles from $^8$B decay were detected in an uncollimated, close geometry with a ``small" $\approx$139 mm$^2$ 20$\mu$m Si surface-barrier detector, that we label ``S", and, in separate measurements, with a ``large" $\approx$416mm$^2$ 33$\mu$m detector, that we label ``L". We inferred the solid angle $\Omega$ of each of these detectors from the counting rate of a custom-made $^{148}$Gd $\alpha$-source deposited on a Mo target backing of the same design as was used for the $^7$Be target.   The source activity was calibrated with a $\sim$450 mm$^2$ 35$\mu$ detector collimated to an area of 246.0 $\pm$ 0.3 mm$^2$, and a source-to-collimator distance of 57.00 $\pm$ 0.10 mm, corresponding to a geometrical solid angle $\Omega_{\rm cal}$ = 0.0744 $\pm$ 0.0003 sr.  The efficiency of the calibration detector for events lying within the collimator acceptance was checked by measurements with different size collimators and different detectors.  From the measured $\alpha$-source counting rates we determined $\Omega/\Omega_{\rm cal}$ with a statistical precision of $\pm$ 0.2\%.  We applied (0.6 $\pm$ 0.6)\% and (0.3 $\pm$ 0.3)\% corrections to the deduced solid angles for the S and L detectors, respectively, to account for a measured (0.03 $\pm$ 0.03) mm difference in the distance from the detector to the BE3 target and to the $^{148}$Gd source.  The results are $\Omega/\Omega_{\rm cal}$ =  24.17 $\pm$ 0.39 and 18.78 $\pm$ 0.13, and hence $\Omega$ =  1.798 $\pm$ 0.030 sr and 1.397 $\pm$ 0.011 sr for the S and L detectors, respectively.

To understand better the accuracy of these solid-angle measurements, we measured the L-detector effective solid angle
as a function of distance using a precision translation stage.  Several different sources, including commercially-produced $^{148}$Gd and $^{241}$Am sources were employed.  Based on the observed spread of these measurements, we conservatively assigned an additional scale-factor (common mode) error of $\pm$ 1.5\% to our determinations of $\Omega$ for both the S and L detectors (see Table \ref{tableS17errors}).

This method had several improvements over the BE1 experiment~\cite{junghans}.  1) It eliminated the $^7$Li(d,p)$^8$Li solid-angle calibration which depended on calculated corrections for the portion of the $^8$Li $\alpha$-spectrum lying below the experimental threshold.
2) The smaller $\alpha$-detector solid angles helped reduce the solid-angle uncertainty.   3) The thinner detector minimized corrections for the portion of the $^8$B $\alpha$-spectra lying below the experimental threshold.

\subsection{Beam-energy calibration}

The accelerator-energy calibration in the BE1 experiment was determined by measuring the $^{19}$F(p,$\alpha \gamma)^{16}$O  resonances
at  340.46 $\pm$ 0.04, 483.91 $\pm$0.10 and 872.11 $\pm$ 0.20 keV~\cite{f19pag} using a thick LiF target.  The measured thick-target yield curves were
fitted with the integral of a Lorentzian folded with a Gaussian beam energy resolution.   Multiple measurements of each resonance, in which the beam tune and steering were varied, were used to estimate the systematic error.

\begin{figure}
\includegraphics[width=0.5\textwidth]{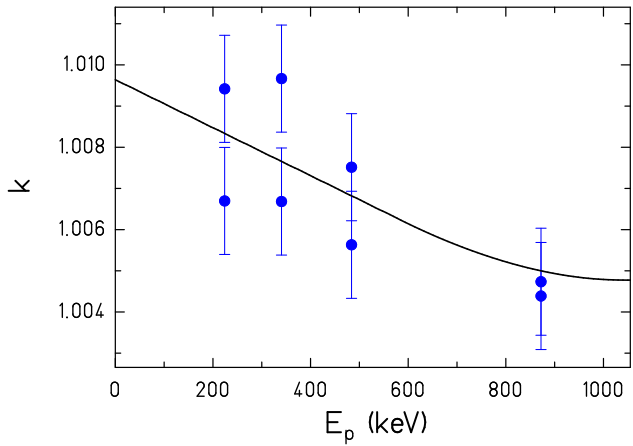}
\caption{Accelerator energy calibration constant $k$ vs. proton energy.  Points: measured values; solid line: spline fit.  The error bars were determined from their scatter relative to the fit.  The $^7$Be($\alpha, \gamma)^{11}$C resonance measurement (at an equivalent
proton energy of 5471 keV) indicated that $k$ was constant above E$_p$ = 1000 keV.}
\label{accelcalib}
\end{figure}

In the BE3 experiment, we remeasured these same resonances, as well as the $^{19}$F(p,$\alpha \gamma)^{16}$O resonance at  $E_p$ = 223.99 $\pm$ 0.07 keV ,  and the $^7$Be($\alpha, \gamma)^{11}$C resonance which we take to be located at 1377 $\pm$ 2 keV from the mean of our resonance energy determination in the BE1 experiment and the determination given in ref. \cite{be7ag}.   A spline curve was fitted through nine measurements of these five resonances and used to determine the accelerator calibration constant $k$, shown in Fig.~\ref{accelcalib}, where
\begin{equation}
E_p = \frac {kf^2}{ 1 + E_p/2m_pc^2}
\label{accelcalequation}
\end{equation}
and $f$ is the magnet NMR frequency.  The BE2 and BE3 measurements were made under similar accelerator conditions, and hence we used the same accelerator calibration for both of these experiments.

\subsection{Alpha spectra from $^7$Be(p,$\gamma$)$^8$B}

\begin{figure}
\includegraphics[width=0.45\textwidth]{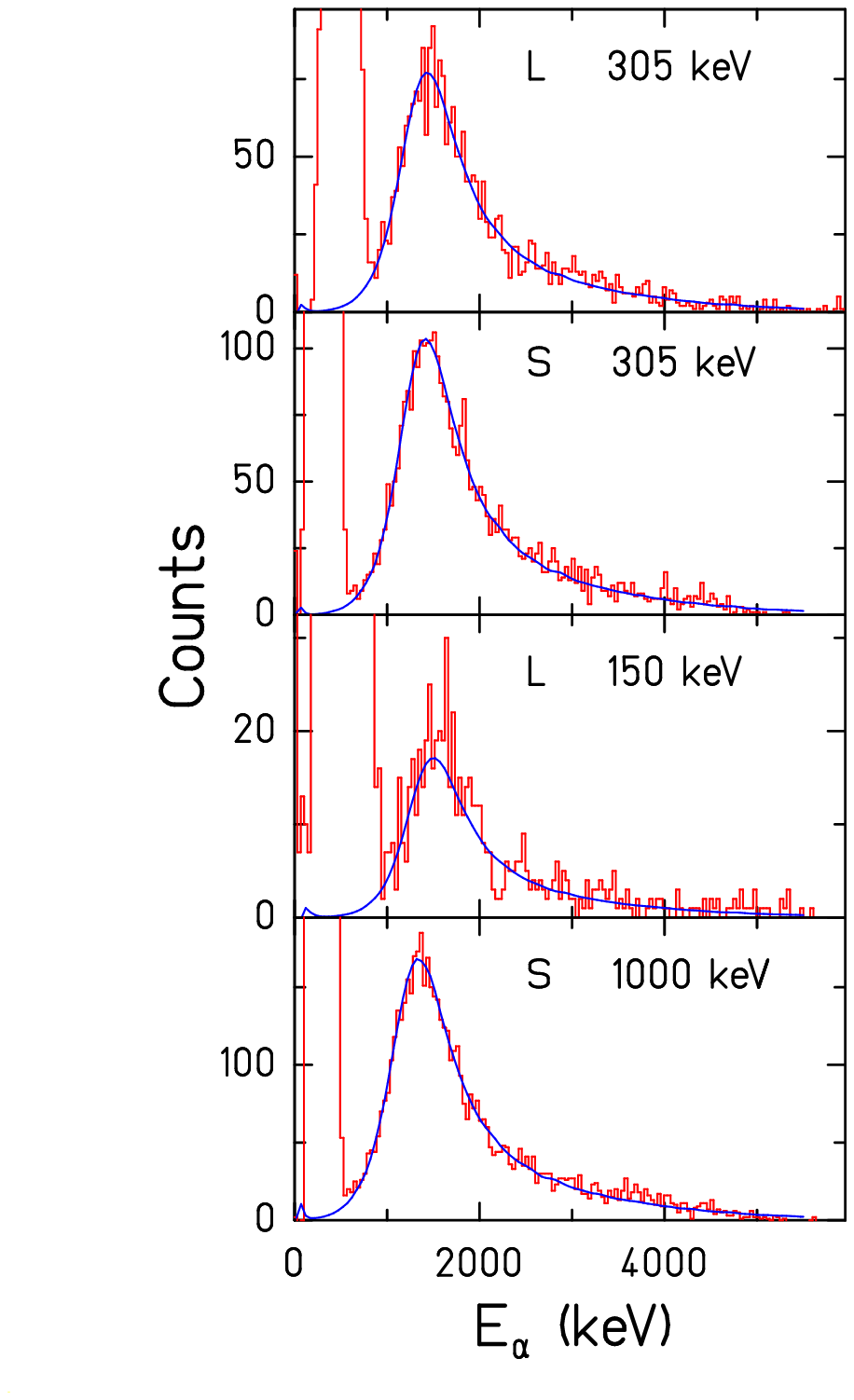}
\caption{Alpha spectra measured with the BE3 target and the S and L detectors, at different proton energies, as indicated.  The curves are fitted TRIM calculations. }
\label{alphaspectra}
\end{figure}

Fig.~\ref{alphaspectra} shows representative $\alpha$-spectra measured with the BE3 target.  The energy scale was determined using a $^{148}$Gd source to monitor the detector gain and a precision pulser to monitor the zero.  Both gain and zero were checked frequently during the experiment, and found to be stable, the gain to $\pm$ 0.2\% and the zero to $\pm$ 3 keV.
At low energies the $\alpha$-spectrum was obscured by background from pileup of Compton electrons produced by the intense 478 keV radiation from the target, and at even lower energies by electronic noise.   The $\alpha$-spectra were summed above a software threshold of 650 keV (900 - 930 keV) in the S (L) detector, and corrected for beam-off background  of  0.6 $\pm$ 0.1 counts/hr (averaged between the S and L detectors), due primarily to $\alpha$-radioactivity in the chamber materials.  A stainless-steel shield between the chamber wall and the Si detector reduced this background by  a factor of 2 relative to the (small) level observed in the BE1 experiment.  This background correction was 5\%, 2\% and 1\% at the 3 lowest proton energies, and less than 0.3\% at higher $E_p$.  Beam-on background was checked at several energies and found to be negligible.  Separate measurements with a LiF target ensured that any deuteron beam contamination (that could contaminate the $\beta$-delayed $\alpha$-spectrum by the $^7$Li(d,p)$^8$Li reaction) was negligible.

To compute the total yield, the data integrated as described above must be corrected for the fraction $F_{\alpha}(E_p)^{-1}$ of the spectrum that lies below the detector threshold.
This was done using the TRIM Monte Carlo (MC) code \cite{srim} to model the implantation depth of the $^8$B ions and the energy loss of the emitted $\alpha$'s.  The $\alpha$-energy spectrum was taken from the thin-target spectrum of ref.~\cite{warburton}.  Calculations were performed in which a) the opening angle of the emitted $\alpha$'s was restricted to the geometrical acceptance of the detector, and b) the $\alpha$'s were randomly emitted into 2$\pi$.  In case b), which allows for multiple scattering into and out of the detector acceptance, additional events were found at low $E_{\alpha} \lesssim $ 200 keV due to large-angle multiple scattering.  However, the $F_{\alpha}(E_p)$ values predicted by the 2 methods
agreed within the MC precisions of $\pm$ 0.2\%.  The curves shown in Fig.~\ref{alphaspectra} were computed with method b), for which
\begin{equation}
F_{\alpha}(E_p) = \frac{\Omega/2\pi }{  \int_{th}^{\infty}N_{\alpha}(E_{\alpha})dE_{\alpha}/N_{tot}},
\label{cf}
\end{equation}
where $\Omega$ is the geometrical solid angle, $\int_{th}^{\infty}N_{\alpha}(E_{\alpha})dE_{\alpha}$ is the integral of all MC events above the detector threshold, and $N_{tot}$ is the total number of MC events.

The S detector spectra (except for the highest energy point at $\bar{E}_{\rm cm}$ = 1754 keV) were fitted with a fixed energy calibration, varying only
the counting-rate normalization.
For the 1754 keV data point and for most of the L-detector data, we had to vary the zero offset to fit the low-energy side of the spectra.  The resulting fitted offsets (typically 20 - 30 keV  with a maximum of 88 keV) are larger than can be accounted for by the measured offsets.  They are also larger than the  1 - 2 keV expected, on average, from $\beta$ - $\alpha$ summing, and suggest a failure of the TRIM calculation.   A similar problem occured  in comparisons of TRIM calculations with measured $^7$Li(d,p)$^8$Li spectra.  As a result of these difficulties, and also problems in the BE2 experiment understanding the ratio of $^7$Li(d,p)$^8$Li yields to $\alpha$-source count-rates for different detector solid angles, we assigned a conservative uncertainty of $\pm$ 30\% on the correction $1 - F_{\alpha}(E_p)$.

For the S (L) detector, $F_{\alpha}(E_p)$ ranged from 1.004  to 1.006  (1.023  to 1.037 ) below the resonance,  and   1.012  to 1.019  (1.054)  above the resonance .

Threshold correction factors for the BE2 experiment are discussed in Sec.~\ref{be1results} below.

\subsection{Backscattering measurements}
\label{backscattering}
Weissman et al.~\cite{weissman} pointed out that previous $^7$Be(p,$\gamma$)$^8$B experiments suffered from unknown losses of $^8$B due to backscattering out of the target, and $^8$Li losses when $^7$Li(d,p)$^8$Li was used for absolute cross section normalization.  Substantial backscattering losses may occur when a high-Z target backing is used or if there are high-Z contaminants in the target.  Proton backscattering followed by $^8$B production is not important.

\begin{figure}
\includegraphics[width=0.45\textwidth]{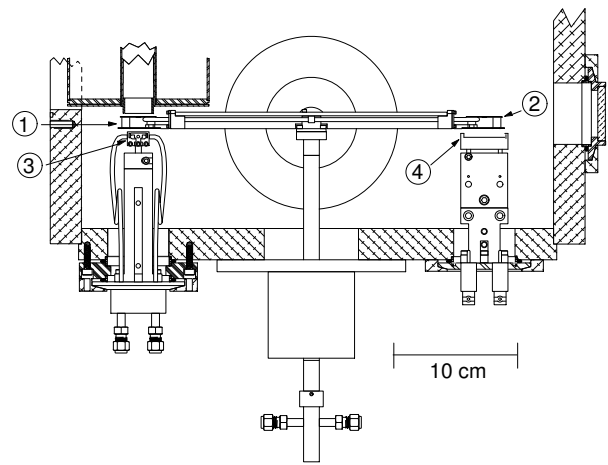}
\caption{Backscattering apparatus.  1 and 2: catcher plates; 3:  fixed target and water-cooled mount; 4: $\alpha$-detector.}
\label{backscattdiag}
\end{figure}

We have made the only measurements of $^8$B backscattering in the $^7$Be(p,$\gamma$)$^8$B reaction, by modifying our apparatus as shown in Fig.~\ref{backscattdiag}.
We installed the $^7$Be target in a fixed mount in place of the Faraday cup, mounted large-area Cu catcher plates on both ends of the rotating arm, and installed the L-detector on the downstream side of the arm.  During target bombardment, the proton beam passed through a 4 mm aperture in the center of the catcher plate before striking the target, and backscattered $^8$B's were caught on the catcher plate.  The arm-rotation time sequence was the same as in the $^7$Be(p,$\gamma$)$^8$B measurement.  The arm and the target were both biased to +300V.  Because of secondary electron crosstalk between the target and the arm, only the total beam current could be measured reliably in this setup.  Hence beam tuning and beam transmission measurements through the 4 mm aperture were done with the Faraday cup in place, after which the $^7$Be target was installed and the backscattering measurements were carried out.  The fraction of the beam passing through the aperture was checked at the end of each measurement by re-installing the Faraday cup.
\begin{figure}
\includegraphics[width=0.5\textwidth]{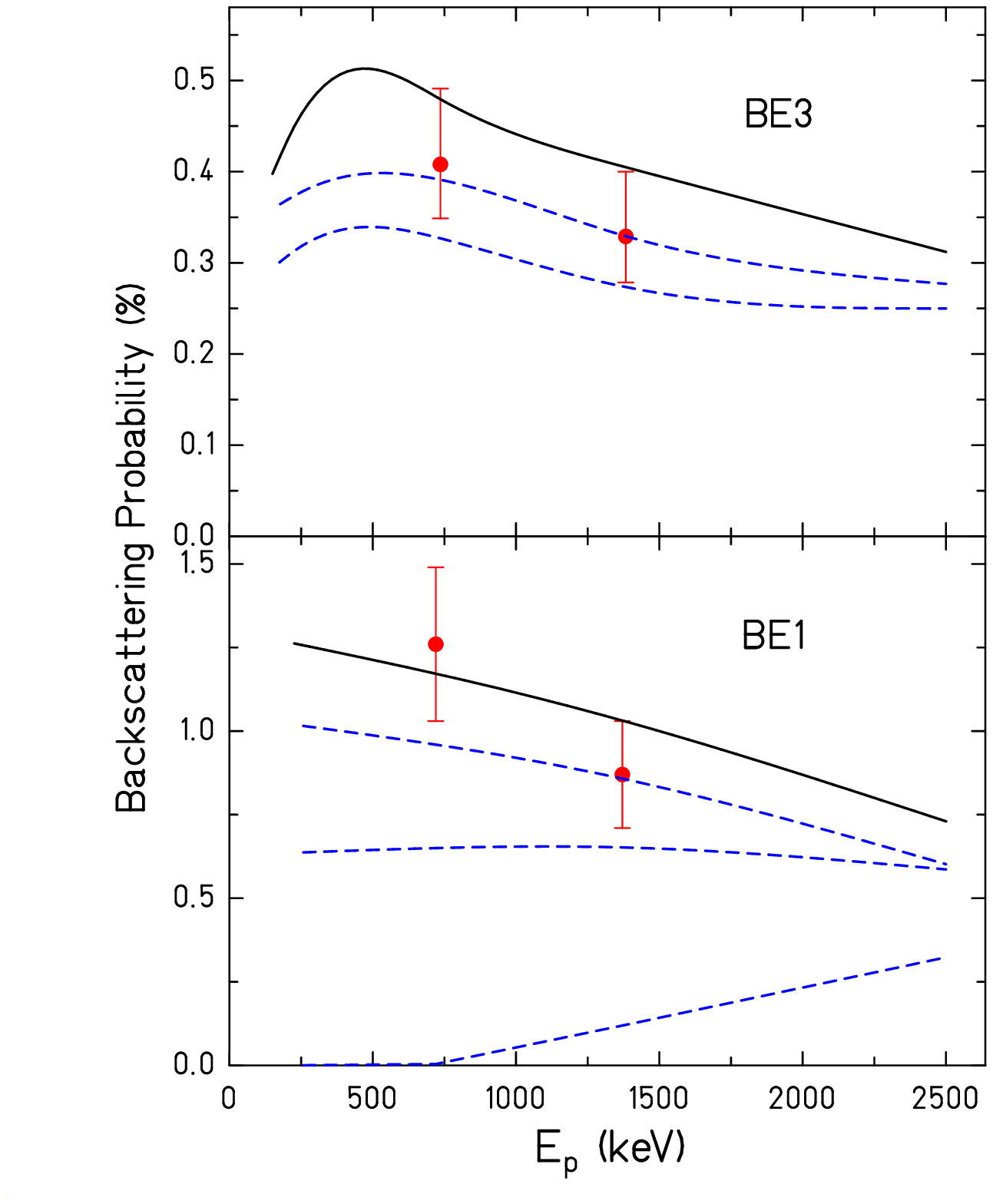}
\caption{Measured backscatting probabilites (errors include statistics and systematics) and TRIM calculations. The $^7$Be:C:Mo atom-number ratios assumed in the TRIM calculations are, in descending order: top panel - 63:0:37, 58:8:34, 57:13:30; bottom panel - 42:0:58, 38:19:43, 36:31:33 and 19:81:0.}

\label{backscattresults}
\end{figure}
Backscattering measurements were made at $\bar{E}_{\rm cm}$ = 626 and 1200 keV.   The efficiency for catching the $^8$B on the plates and counting the subsequent $\beta$-delayed $\alpha$-particles was computed using TRIM to estimate the backscattered $^8$B angular distribution. Figure~\ref{backscattresults} shows the measured backscattering probabilities for the  BE1 and BE3 targets, along with  TRIM calculations
for several assumed target compositions.
We assumed uniform targets composed of $^7$Be, C and Mo, where C is representative of low-Z contaminants and Mo is representative of high-Z contaminants (see Sec.\ \ref{alphagamma}).  The amount of $^7$Be in the target is fixed by the target activity, and the amount of contaminants is fixed by the ($\alpha,\gamma$) resonance profile width in excess of the width expected for pure $^7$Be (see Sec.~\ref{alphagamma}).

The qualitative shapes of these curves are easy to understand.  For pure C contaminant, the backscattering is primarily from the higher-Z Mo backing.  Although the (Rutherford) backscattering cross section rises as the energy drops, at low bombarding energy the backscattered $^8$B ions have insufficient energy to escape from the target layer.  For pure Mo contaminant, the backscattering probability is higher and extends to lower energies, since the backscattering may now occur in the target layer.  The backscattering probability from BE3 is smaller than from BE1 because this target was thicker which suppressed backscattering from the Mo backing at low energies.

As shown in Fig.~\ref{backscattresults}, the best fit curves are $^7$Be:C:Mo = 38:19:43 for BE1, and  $^7$Be:C:Mo = 58:8:34 for BE3.  These contaminant compositions are similar, which is expected since the target fabrication process was the same.  We note that a pure Mo contaminant is consistent with both target compositions, within errors.  The backscattering probability can be affected by target nonuniformity.  Even though BE1 was significantly less uniform than BE3, as can be seen from the $^7$Be($\alpha, \gamma)^{11}$C profiles, the agreement between measured and calculated backscattering probabilities with similar contaminant compositions suggests that nonuniformity did not play an important role.

Based on Fig.~\ref{backscattresults}, we made constant backscattering corrections to our measured cross sections of 0.4 $\pm$ 0.1\% for BE3 and 1.0 $\pm$ 0.5\% for BE1.

\section{BE3 results}
\label{be3results}

\subsection{Data}
\label{subsection: data}

\begin{figure}
\includegraphics[width=0.5\textwidth]{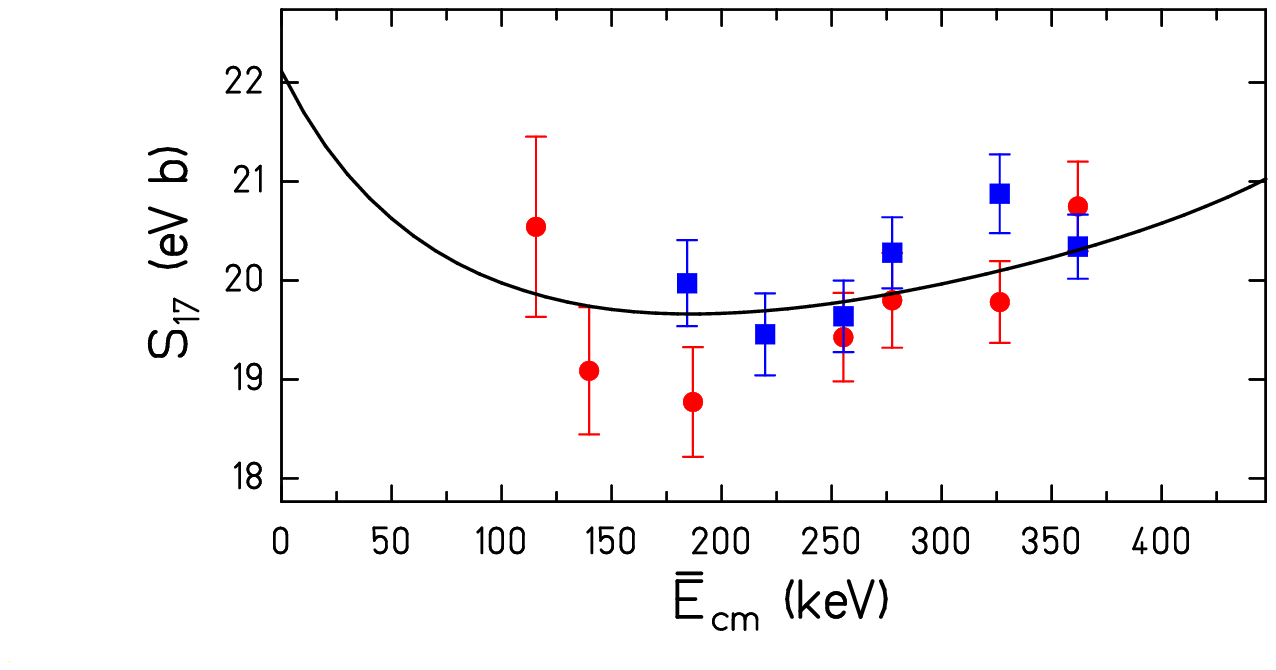}
\caption{BE3 S (circles) and L (squares) S-factor data measured below the resonance.  Curve: best DB fit.  Error bars include statistical and varying systematic errors.}
\label{SandL}
\end{figure}

Figure~\ref{SandL} shows the data from the BE3 experiment.  The cross sections were converted into S-factors using the relation
\begin{equation}
S_{17}(\bar{E}_{\rm cm}) = \sigma(\bar{E}_{\rm cm})\bar{E}_{\rm cm}
e^{(E_G/\bar{E}_{\rm cm})^{1/2}}
\label{Sfactor}
\end{equation}
where the Gamow energy $E_G = (2\pi \alpha Z_1 Z_2)^2 \mu c^2/2$ = 13799.3 keV,
and $\bar{E}_{\rm cm}$ values were computed as described below.  This procedure corrects for energy averaging, so that fitting experimental $S_{17}(\bar{E}_{\rm cm})$ values without explicitly including energy averaging is equivalent to fitting the measured cross sections including energy averaging.
Table~\ref{tableS17errors} lists the uncertainties in the experimental S$_{17}$(E) values.

\subsection{$\bar{E}_{\rm cm}$ values}

We computed the effective energy $\bar{E}_{\rm cm}$ of the data points as follows.  First, we computed $\langle\sigma (E_{\rm cm})\rangle$  using the appropriate $^7$Be($\alpha, \gamma)^{11}$C profile (see Sec.\ \ref{alphagamma}),   where  $\sigma (E_{\rm cm})$ is given by Eq.\ (\ref{Sfactor}) with $S_{17}(E_{\rm cm})$ set equal to a constant.  Here we assumed a pure Mo contaminant which is consistent with
the backscattering data -- see Sec.\ \ref{backscattering}. Then we solved the equation  $\langle\sigma (E_{\rm cm})\rangle$ = $\sigma (\bar{E}_{\rm cm})$ for $\bar{E}_{\rm cm}$.  Our neglect of the variation of $S_{17}(E_{\rm cm})$ over the target thickness is a good approximation everywhere except near the resonance.  Our analysis of the BE1 data, which included the resonance, took into account
the effect of the resonance on the calculated $\bar{E}_{\rm cm}$'s -- see Sec.\ \ref{be1results}.  Away from the resonance, $\bar{E}_{\rm cm}$ and
$\langle E_{\rm cm}\rangle$ values are similar everywhere except at the lowest energies; for example, at E$_p$ = 149.9 keV, the energies $\bar{E}_{\rm cm}$ = 113.9 keV and $\langle E_{\rm cm}\rangle$ = 115.6 keV correspond to a 6\% difference in $S_{17}(E_{\rm cm})$.

\begin{table}
\caption{Percent uncertainties in S$_{17}$ from the renormalized BE3 (S and L) data.}
\label{tableS17errors}
\begin{ruledtabular}
\begin{tabular}{ld}
Statistical errors & 1.3-4.0\\
    \hline
Varying systematic errors: &\\
\hspace{0.5cm}  proton energy calibration & 0.0-0.7\\
\hspace{0.5cm}  target thickness & 0.0-1.6\\
\hspace{0.5cm}  target composition & 0.0-0.7\\
\hspace{0.5cm}  $\alpha$-spectrum cutoff & 0.1-0.7 (S)\\
\hspace{0.5cm}  & 1.0-1.8 (L)\\
   \hline
Scale factor errors: &\\
\hspace{0.5cm}  beam-target inhomogeneity & 1.0\\
\hspace{0.5cm}  integrated beam flux & 0.9\\
\hspace{0.5cm}  $^7$Be target atom number & 1.1\\
\hspace{0.5cm}  solid angle & 1.5\\
\hspace{0.5cm}  backscattering & 0.1\\
\hspace{0.5cm}  timing cycle & 0.2\\
Total scale factor error & 2.3\\
\end{tabular}
\end{ruledtabular}
\end{table}

\subsection{Determination of S$_{17}(0)$}
Because the experimental data must be extrapolated to low energy to determine the astrophysical S-factor, it is best to fit data as low in energy as possible, commensurate with good experimental precision.
Above the M1 resonance at $\bar{E}_{\rm cm}$ = 630 keV, different $^7$Be(p,$\gamma)^8$B cross-section calculations deviate substantially from one another.  Below the resonance, the $^7$Be(p,$\gamma)^8$B  reaction is predominantly direct capture, and becomes increasingly extranuclear and hence less model-dependent with decreasing $E_p$.

\begin{figure}
\includegraphics[width=0.5\textwidth]{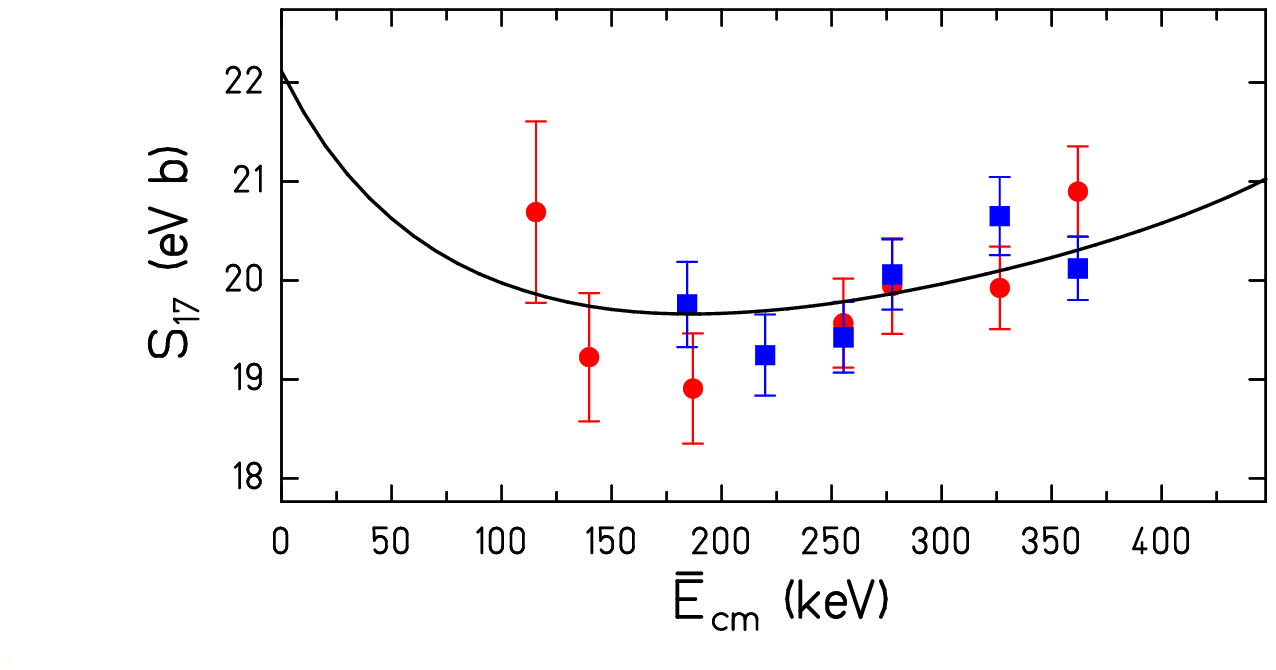}
\caption{BE3 S and L S-factor data renormalized to the best-fit S$_{17}(0)$ = 22.1 eV b, otherwise as in Fig.~\ref{SandL}.}
\label{SandLrenormalized}
\end{figure}

The $^7$Be(p,$\gamma)^8$B calculation that fits experimental data best over a wide range is the cluster-model theory of Descouvemont and Baye~\cite{db} (DB) (see Sec.~\ref{extrapolation} below).  Figure~\ref{SandL} shows our fit of the scaled DB theory to all our data with $\bar{E}_{\rm cm} $  = 116 to 362 keV.  The DB theory shown here does not contain contributions from either the 1$^+$ resonance near 630 keV or the 3$^+$ resonance near 2200 keV.  For these low-energy data the contribution of the 1$^+$ resonance is less than 0.4\% and the contribution of the 3$^+$ resonance is completely negligible based on our resonance fits discussed below.  The resulting S$_{17}(0)$ values are 22.35 $\pm$ 0.41 eV b  and 21.95 $\pm$ 0.34 eV b for the S and L detector data, respectively.  These values agree well within the quoted errors, which include only statistical and relative (independent) systematic contributions.    Our best value for S$_{17}(0)$ is the weighted average of these 2 results.  Including the common-mode scale-factor error of $\pm$ 2.3\% from Table~\ref{tableS17errors}, we obtain
\begin{equation}
S_{17}(0) = 22.1 \pm 0.6 \mbox{(expt)} \hspace{0.2cm} \mbox{eV b~,}
\label{bestSfactor}
\end{equation}
where the error includes all contributions other than the theoretical extrapolation uncertainty discussed below.
The curve shown in Fig.~\ref{SandL} corresponds to the DB calculation scaled to our best fit value for
S$_{17}(0)$ as given in Eq.\ (\ref{bestSfactor}) above.   Fig.~\ref{SandLrenormalized} shows these same data renormalized to the best fit value for S$_{17}(0)$.

\section{BE1 and BE2 results}
\label{be1results}
\begin{figure}
\includegraphics[width=0.5\textwidth]{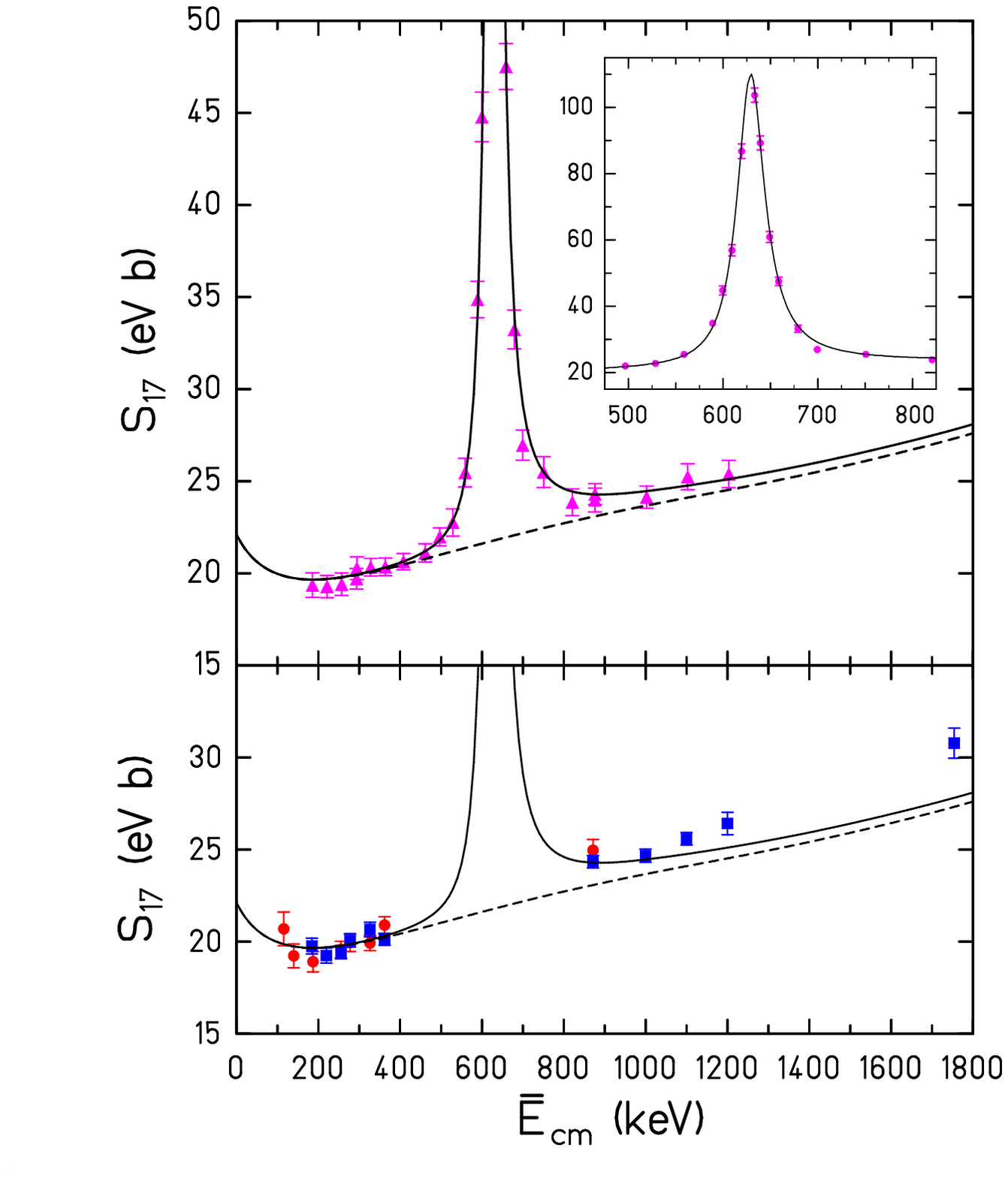}
\caption{Top panel: BE1 data normalized to S$_{17}$(0) = 22.1 eV b.  Solid curve: best-fit DB plus a fitted 1$^+$ resonance; dashed curve - DB only (see text).  Inset: resonance region.  Bottom panel: solid squares - BE3 S data; solid circles - BE3 L data.  The solid curve was calculated with 1$^+$ resonance parameters determined from fits to the BE1 data, and the normalization was determined by fitting the BE3 data with $\bar{E}_{\rm cm}$ $\leq$ 362 keV.}
\label{BE1andBE3Sfactor}
\end{figure}

Results for our BE1 experiment were reported in ref.~\cite{junghans}.   Figure~\ref{BE1andBE3Sfactor} and Table~\ref{allourresults} show these results, renormalized to our best value for S$_{17}(0)$ as given in Eq.\ (\ref{bestSfactor}).  The error bars have been increased relative to those shown in Fig.3 of ref.~\cite{junghans}, because of the increased $F_{\alpha}(E_p)$ uncertainty discussed above; otherwise, the data are the same as in
ref.~\cite{junghans}.
Note that our previously published value, $S_{17}(0)$ = 22.3 $\pm$ 0.7(expt) eV b~\cite{junghans}, is in excellent agreement with our new value presented above.

\begin{figure}
\includegraphics[width=0.5\textwidth]{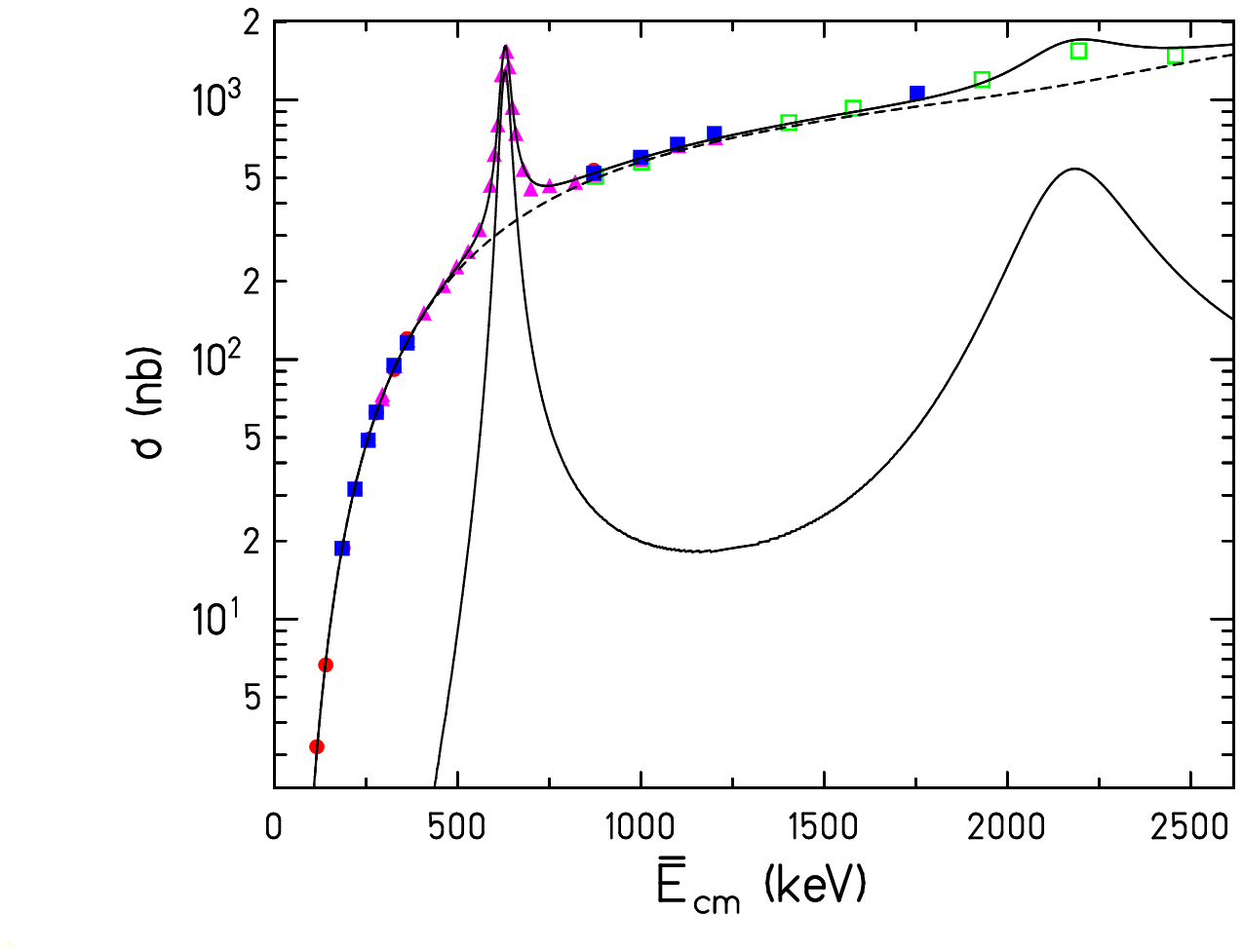}
\caption{Cross-sections from the BE1, BE2 and BE3 experiments.  The BE2 data are shown as open squares, and the BE1 and BE3 data are shown with the same symbols as in Fig.~\protect\ref{BE1andBE3Sfactor}.  Solid curve: best-fit DB plus fitted 1$^+$ and 3$^+$ resonances; dashed curve: DB only; lower solid curve: 1$^+$ and 3$^+$ resonance contributions; as in Fig~\protect\ref{BE1andBE3Sfactor}.}
\label{crosssection}
\end{figure}

\begin{figure}
\includegraphics[width=0.5\textwidth]{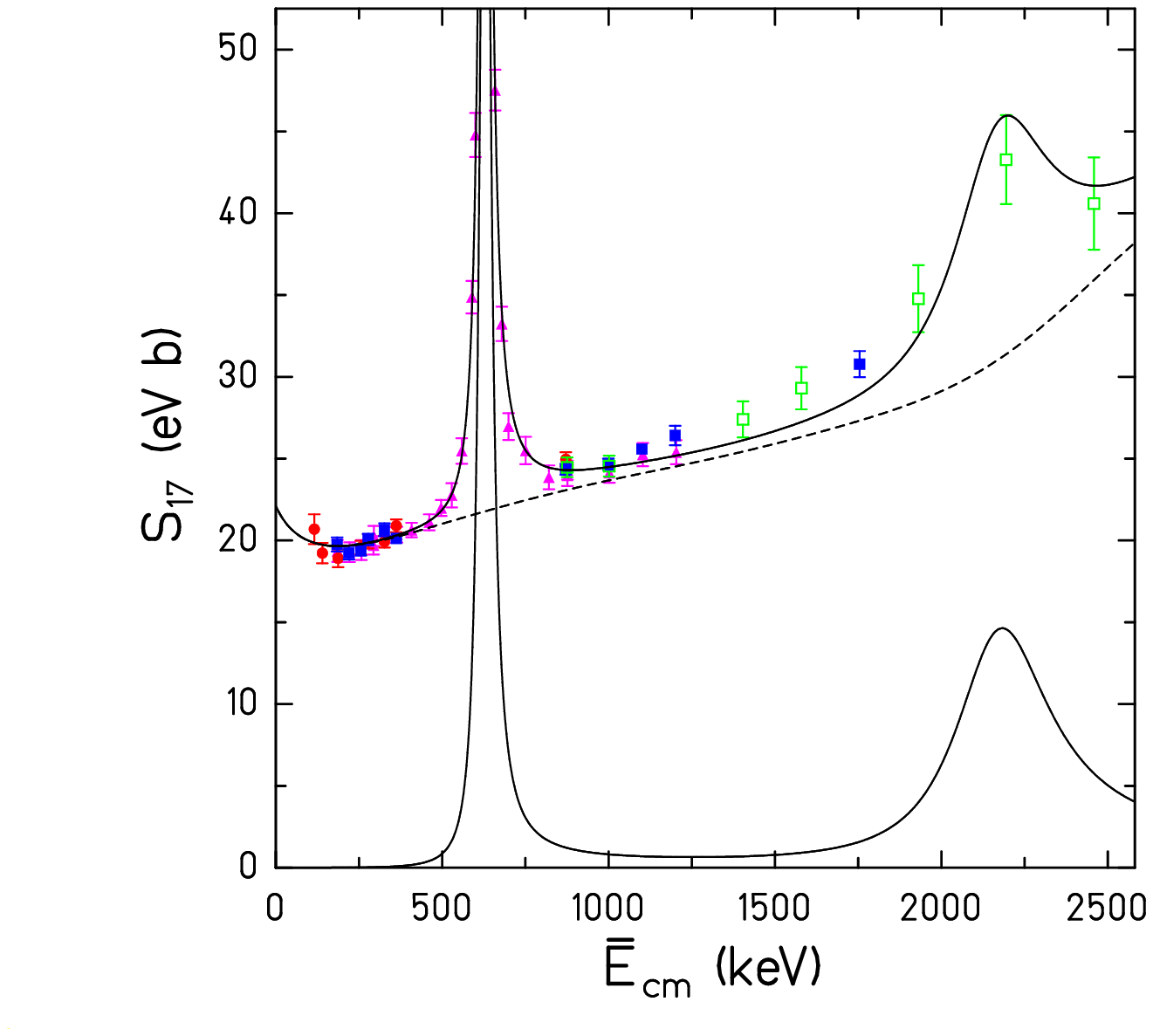}
\caption{Measured and calculated S-factors corresponding to the cross sections shown in the previous figure.   Solid curve: best-fit DB plus fitted 1$^+$ and 3$^+$ resonances; dashed curve: DB only; lower solid curve: 1$^+$ and 3$^+$ resonance contributions.}
\label{allourSfactordata}
\end{figure}

In the BE2 experiment, data were taken in a close geometry with the L detector, over the range $\bar{E}_{\rm cm}$ = 876 to 2459 keV.
These data are shown in Fig.~\ref{crosssection} and ~\ref{allourSfactordata} where they are
plotted, along with our BE1 and BE3 data with a common absolute normalization determined by our BE3 results (Eq.~\ref{bestSfactor2}).  The BE2 data
have relatively large systematic errors due to the large threshold correction factors $F_{\alpha}(E_p)$ that range from 1.091  to 1.287, and the corresponding $\pm$ 30\% uncertainty on  $1 - F_{\alpha}(E_p) $.

\subsection{$1^+$ resonance}
\label{m1resonance}
In the BE1 experiment, data were taken from $\bar{E}_{\rm cm}$ = 186 to 1200 keV, including detailed measurements over the 630 keV M1 resonance.  $\bar{E}_{\rm cm}$ values were determined by fitting the cross-section data, including averaging over the target profile, with the sum of the nonresonant DB cross section plus an incoherent Breit-Wigner resonance (see below), and then solving the equation $\bar{\sigma} = \sigma(\bar{E}_{\rm cm})$ for $\bar{E}_{\rm cm}$, where $\bar{\sigma}$ is the measured
cross section and $\sigma(\bar{E}_{\rm cm})$ is the fitted unaveraged cross section at a given proton bombarding energy.  This procedure removes the effects of energy averaging, as discussed above.  Measured cross sections were converted to S-factors using Eq.\ (\ref{Sfactor}).   We checked that fitting our S$_{17}( \bar{E}_{\rm cm})$ data without energy averaging, gave the same parameters as the original fit to the cross section including energy averaging.

The cross section was fitted with
\begin{equation}
\sigma(\bar{E}_{\rm cm}) = C_1 \sigma_{DB}(\bar{E}_{\rm cm}) + \frac{C_2 }{ \bar{E}_{\rm cm}}\frac{\Gamma_p(\bar{E}_{\rm cm})\Gamma_{\gamma}(\bar{E}_{\rm cm}) }{ (\bar{E}_{\rm cm} - E_0)^2 + \Gamma_p(\bar{E}_{\rm cm})^2/4}
\label{sigmafit}
\end{equation}
where $C_1$ $\approx$ 0.7 is a fitted scaling factor,  $\sigma_{DB}(\bar{E}_{\rm cm})$ is the DB cross section (with the $1^+$ resonance removed),  $C_2 = 3 \pi \lambdabar^2 \bar{E}_{\rm cm}/8$ , $\Gamma_p(\bar{E}_{\rm cm}) = \Gamma_p(E_0)P_1(\bar{E}_{\rm cm})/P_1(E_0)$, $P_1(\bar{E}_{\rm cm})$ is the $\ell$=1 Coulomb penetrability evaluated at R = 3.65 fm, $\Gamma_{\gamma}(\bar{E}_{\rm cm}) = \Gamma_{\gamma}(E_0)E_{\gamma}^3/E_0^3$,  $E_{\gamma}$ = $\bar{E}_{\rm cm}$ + $Q$ and $Q$ = 0.137 MeV.

Table~\ref{tablerespar} shows our center-of-mass 1$^+$ resonance fit parameters together with those of refs.~\cite{filippone,baby} and  the recent elastic scattering results of ref.~\cite{angulo}.  Descouvemont and Baye~\cite{db} predict the lowest 1$^+$ resonance at E$_{\rm cm} \sim$ 0.2 MeV.  Scaling by the experimentally measured energy, they calculate  $\Gamma_p(E_0)$ $\approx$ 59 keV and  $\Gamma_{\gamma}(E_0)$ = 33 meV (assuming pure M1), in reasonable agreement with experiment.
\begin{table}[h]
\caption{1$^+$ resonance parameters}
\label{tablerespar}
\begin{ruledtabular}
\begin{tabular}{lcccc}
Parameter & Present work &  Ref.~\protect\cite{filippone}  &Ref.~\protect\cite{baby} & Ref.~\protect\cite{angulo} \\
    \hline
$E_0$ (keV) & 630 $\pm$ 3 & 632 $\pm$ 10 & 633 & 634 $\pm$ 4 \\
$\Gamma_p(E_0)$ (keV)  & 35.7 $\pm$ 0.6 & 37 $\pm$ 5 & 35 $\pm$ 3 & 31 $\pm$ 4\\
$\Gamma_{\gamma}(E_0)$ (meV) & 25.3 $\pm$ 1.2 & 25 $\pm$ 4 & 25 $\pm$ 2 & - \\
\end{tabular}
\end{ruledtabular}
\end{table}

\subsection{3$^+$ resonance}
\label{3resonance}

Our data in  Fig.~\ref{allourSfactordata} show clear evidence for the lowest 3$^+$ resonance at E$_{\rm cm}$ $\sim$ 2200 keV. The DB calculation shown in Figs. \ref{SandL} - \ref{BE1andBE3Sfactor} does not include this 3$^+$ resonance.   To understand better the $^7$Be(p,$\gamma$)$^8$B cross section above $\bar{E}_{\rm cm}$ = 1200 keV, we have fitted our BE2 + BE3 data with the DB calculation plus adjustable 1$^+$ and 3$^+$ resonances.    The 1$^+$ parameters were fixed to our fit results quoted in Table \ \ref{tablerespar} and the 3$^+$ resonance was fitted  with a formula similar to that given in Eq.~(\ref{sigmafit}), in which the constant $C_2$ was multiplied by the factor 7/3 to  account for the $J = 3$ resonance angular momentum.  This formula neglects f-wave capture and $E2$ decay, which should be good approximations.  We also neglect the $^4$He + $^3$He + p channel and (here as in the 1$^+$ resonance analysis) proton inelastic scattering to the first excited state of $^7$Be.

The fit results are shown in Table \ref{3+table}.  The unconstrained fit parameters, $E_0$ = 2100 $\pm$ 60 keV, $\Gamma_p(E_0)$ = 510 $\pm$ 270 keV and  $\Gamma_{\gamma}(E_0)$ = 180 $\pm$ 70 meV, are not well determined. The resonance energy and width agree with the more precise values 2183 $\pm$ 30 keV and 350 $\pm$ 40 keV, respectively, compiled in \cite{ajzenberg}.  Constraining the resonance energy and width to the values from \cite{ajzenberg} results in  the fits shown in Figs.~\ref{crosssection} and \ref{allourSfactordata}, for which $\Gamma_{\gamma}(E_0)$ = 150 $\pm$ 30 meV.  Descouvemont and Baye~\cite{db} also calculated this 3$^+$ resonance, and found $E_0$ $\sim$ 2800 keV.  After adjusting their resonance energy to agree with experiment, they obtained  $\Gamma_p(E_0)$ = 530  keV and  $\Gamma_{\gamma}(E_0)$ = 45  meV.

\begin{table}
\caption{3$^+$ resonance parameters}
\label{3+table}
\begin{ruledtabular}
\begin{tabular}{lcc}
Parameter & Constrained fit &  Unconstrained fit  \\
    \hline
$E_0$ (keV) & 2183\footnotemark[1] & 2100 $\pm$ 60  \\
$\Gamma_p(E_0)$ (keV)  & 350\footnotemark[1] & 510 $\pm$ 270 \\
$\Gamma_{\gamma}(E_0)$ (meV) & 150 $\pm$ 30 & 180 $\pm$ 70  \\
\end{tabular}
\end{ruledtabular}
\footnotetext[1]{ref.~\protect\cite{ajzenberg}.}
\end{table}

The DB calculation together with fitted 1$^+$ and 3$^+$ resonances provides a reasonable description of the data up to $\bar{E}_{\rm cm}$ = 2500 keV, although the constrained fit indicates that the DB ``background" underneath the 3$^+$ resonance is not quite correct.  Including the fitted $3^+$ resonance has very little ($\leq$ 1\%) effect below 1500 keV; hence we do not include the fitted 3$^+$ resonance elsewhere in this paper.

\begin{table}
\caption{S$_{17}$ data normalized to the best-fit BE3 results, and $1\sigma$ errors from this work. $\sigma_{\rm stat}$ - statistical error,  $\sigma_{\rm vary}$ - varying systematic error, all in eV b.  An additional common-mode error of 2.3\% applies to each point. }
\label{allourresults}
\begin{ruledtabular}
\begin{tabular}{rccc|rccc}
$\bar E_{\rm cm}$ & S$_{17}$ & $\sigma_{\rm stat}$ & $\sigma_{\rm vary}$ & $\bar E_{\rm cm}$ & S$_{17}$ & $\sigma_{\rm stat}$ &
$\sigma_{\rm vary}$\\
   \hline
\multicolumn{4}{c|}{BE3 S}&\multicolumn{4}{c}{BE3 L}\\
   \hline
184.3&19.8&0.4&0.2&115.6&20.6&0.8&0.4\\
219.8&19.3&0.4&0.2&139.8&19.2&0.6&0.3\\
255.4&19.4&0.3&0.1&187.0&18.9&0.5&0.2\\
277.5&20.1&0.3&0.1&255.3&19.5&0.4&0.2\\
326.4&20.7&0.4&0.1&277.5&19.9&0.4&0.2\\
361.9&20.1&0.3&0.1&326.4&19.9&0.3&0.2\\
871.2&24.3&0.3&0.1&361.9&20.9&0.4&0.3\\
999.5&24.6&0.3&0.1&871.4&25.0&0.4&0.4\\
 \cline{5-8}
1099.8&25.6&0.3&0.1&\multicolumn{4}{c}{BE 2}\\
 \cline{5-8}
1200.1&26.4&0.6&0.1&875.7&24.4&0.2&0.5\\
1754.1&30.8&0.8&0.2&1001.6&24.5&0.2&0.5\\
 \cline{1-4}
\multicolumn{4}{c|}{BE1}&1403.6&27.4&0.4&0.6\\
 \cline{1-4}
185.6&19.3&0.6&0.4&1579.4&29.3&0.5&0.6\\
221.3&19.3&0.5&0.3&1931.0&34.8&0.6&0.8\\
257.0&19.4&0.6&0.3&2194.7&43.2&0.8&0.9\\
293.5&19.6&0.5&0.3&2458.5&40.6&0.7&0.9\\
 \cline{5-8}
294.4&20.2&0.6&0.3&\multicolumn{4}{c}{BE1}\\
 \cline{5-8}
328.2&20.3&0.4&0.3&639.4&89.3&1.4&1.7\\
363.8&20.3&0.4&0.3&649.2&60.9&1.2&1.1\\
408.1&20.6&0.3&0.3&658.7&47.5&0.9&0.9\\
461.3&21.1&0.4&0.3&679.1&33.2&0.9&0.6\\
496.7&22.0&0.3&0.4&699.4&27.0&0.7&0.5\\
528.6&22.8&0.6&0.4&750.7&25.5&0.7&0.5\\
558.8&25.4&0.7&0.4&820.7&23.9&0.5&0.5\\
589.0&34.9&0.8&0.6&876.3&24.3&0.2&0.5\\
599.7&44.8&1.1&0.8&876.3&24.0&0.4&0.5\\
609.4&56.9&1.4&1.0&1002.3&24.2&0.2&0.6\\
619.6&86.8&1.6&1.6&1102.8&25.2&0.3&0.6\\
633.3&103.7&1.1&1.9&1203.2&25.4&0.3&0.7\\
\end{tabular}
\end{ruledtabular}
\end{table}

\section{Discussion}

\subsection{Extrapolation uncertainty}
\label{extrapolation}
The DB calculation with an empirically-fitted 1$^+$ resonance provides the best description of the energy dependence of the $^7$Be(p,$\gamma$)$^8$B cross section in the range $\bar{E}_{\rm cm} \leq $ 1200 keV,  although it
underestimates the data slightly in the region near  1200 keV.

\begin{figure}
\includegraphics[width=0.5\textwidth]{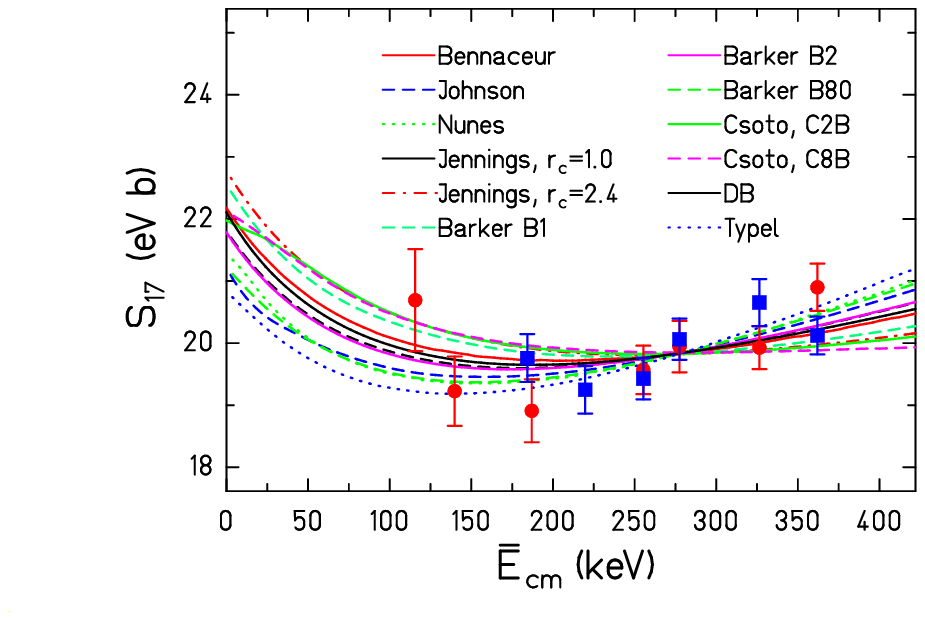}
\caption{Fits of 12 different theories to the BE3 data below the resonance - see \protect\cite{alltheories}.}
\label{extrap}
\end{figure}

It seems reasonable that the DB calculation, which fits the $^7$Be(p,$\gamma$)$^8$B data best over a wide energy range, should be most reliable
for extrapolating to energies of astrophysical interest.
Therefore,  we base our central value for S$_{17}$(0) on the DB extrapolation of our low-energy data.  However, it is  possible that calculations that do less well at high energies may be acceptable at low energies.  We estimated the extrapolation uncertainty by fitting 12 published  calculations~\cite{alltheories} to our BE3 data with $\bar{E}_{\rm cm} \leq $ 400 keV.  The results, shown in Fig.~\ref{extrap}, exhibit a total spread of 2 eV b (out of 22 eV b) at zero energy, and an rms deviation of $\pm$ 0.6 eV b which we adopt as the theoretical extrapolation uncertainty.  Hence our best-fit result is
\begin{equation}
\mbox{S}_{17}(0) = 22.1 \pm 0.6 \mbox{(expt)} \pm 0.6 \mbox{(theor)}
\hspace{0.2cm} \mbox{eV b.}
 \label{bestSfactor2}
\end{equation}

\begin{table}
\caption{$S_{17}(20)$ and $S_{17}(0)$ (in eV b) from fitting our data with $\bar{E}_{\rm cm} \leq $ 362 keV with different models, as in  Fig.~\protect\ref{extrap} and ref.~\protect\cite{alltheories}.}
\label{tableS0S20}
\begin{ruledtabular}
\begin{tabular}{lcc}
Model &$S_{17}(20)$ &$S_{17}(0)$\\
    \hline
Nunes    &      20.8 & 21.4\\
Johnson      &  20.5 & 21.2\\
Bennaceur    &  21.5 & 22.2\\
Barker B80    &  20.7 & 21.2\\
Barker B1    &  21.8 & 22.6\\
Barker B2    &  21.1 & 21.8\\
Csoto C2B    &  21.7 & 22.0\\
Csoto C8B    &  21.8 & 22.1\\
Jennings $r_c=2.4$ fm &  22.0 & 22.8\\
Jennings $r_c=1.0$ fm &  21.1 & 21.8\\
Typel    &      20.3 & 20.8\\
Descouvemont &  21.4 & 22.1\\
\end{tabular}
\end{ruledtabular}
\end{table}

Our theoretical error estimate of $\pm$ 0.6 eV b is somewhat larger than the value quoted in ref.~\cite{junghans} because we now include the Typel calculation.  It is also considerably larger, and hence more conservative than the $\pm$ 0.2 eV b uncertainty recommended by Jennings~\cite{jennings}.  Note that the theoretical (extrapolation) uncertaintly is as large as the experimental uncertainty so that additional theoretical work to reduce the extrapolation uncertainty would be very valuable.

In solar model calculations, the value of S$_{17}$(0) along with the derivatives~\cite{adelberger} S$_{17}^{\prime}$(0)/S$_{17}$(0) and S$_{17}^{\prime \prime}$(0)/S$_{17}$(0) are used to compute the $^7$Be(p,$\gamma$)$^8$B reaction rate in the sun.  However, Jennings has pointed out~\cite{jennings} that the derivatives vary significantly among the different theories, and also differ from the ``best" values given in ref.~\cite{adelberger}.  He~\cite{jennings} argues that S$_{17}$(20) should be used in solar model calculations, which avoids the need for derivatives since 20 keV is near the center of the Gamow window.  Using S$_{17}$(20) instead of S$_{17}$(0) also avoids the need to extrapolate theoretical cross section calculations to zero energy.

For these reasons, we also quote our best-fit result for the S-factor at 20 keV:
\begin{equation}
\mbox{S}_{17}(20) = 21.4 \pm 0.6 \mbox{(expt)} \pm 0.6 \mbox{(theor)}\hspace{0.2cm} \mbox{eV b.}
 \label{bestS20}
\end{equation}
Table \ \ref{tableS0S20} displays our S$_{17}$(0) and S$_{17}$(20) values obtained from the different theoretical extrapolations.  There is a 3\% variation in the ratio S$_{17}$(20)/S$_{17}$(0) among these theories, which is surprisingly large.

\subsection{Comparison with other direct experiments}
\label{directcompare}
We compare the results of all direct experiments by fitting the DB theory to published data in two different energy ranges: $\bar{E}_{\rm cm} \leq$ 425 keV, and $\bar{E}_{\rm cm} \leq$ 1200 keV.  We made a substantial effort to ensure accuracy in these comparisons by obtaining data from primary sources whenever possible, and by fitting the data ourselves.

In the low-energy range, the 1$^+$ resonance contribution may be neglected (it is $\approx$~1\% of the direct contribution at $\bar{E}_{\rm cm}$ = 425 keV and drops rapidly with decreasing energy), and the theoretical uncertainty is minimized.  The experimental uncertainty due to the $\alpha$-threshold correction is also minimized, and the high energy tail of the 1$^+$ resonance is avoided.
On the other hand, some experiments do not have good precision at low energies (none are as good as the present study), which motivates our
wider-range comparison.  In the wide-range fits we included the 1$^+$ resonance with parameters fixed from the fit to our data and excluded data close to the resonance.
Care was taken to separate common-mode (scale factor) errors from other errors.

\begin{table}
\caption{Experimental S$_{17}(0)$ values and uncertainties in eV b determined by our DB fits to published data. }
\label{otherdirectS17values}
\begin{ruledtabular}
\begin{tabular}{lcccc}
Fit Range    &\multicolumn{2}{c}{$\leq425$ keV } &\multicolumn{2}{c}{$\leq1200$ keV}\\
\hline
Experiment   & Value & Error & Value & Error \\
\hline
Filippone        & 20.7 & 2.5 & 19.4 & 2.2\\
Hammache     & 20.1 & 1.3 & 19.4 & 1.1\\
Hass         &      &     & 20.4 & 1.1\\
Strieder     & 18.8 & 1.8 & 18.1 & 1.6\\
Baby             & 20.8 & 1.3 & 21.2 & 0.7\\
This work    & 22.1 & 0.6 & 22.3 & 0.6\\
\hline
${\leq}$425 keV best fit  & 21.4 & 0.5 &     &\\
${\leq}$1200 keV best fit &      &     &21.1 & 0.4\\
\end{tabular}
\end{ruledtabular}
\end{table}

\begin{table}
\caption{Experimental S$_{17}(20)$ values and uncertainties in eV b determined by our DB fits to published data, otherwise the same as Table~\protect\ref{otherdirectS17values}.   }
\label{otherdirectS17(20)values}
\begin{ruledtabular}
\begin{tabular}{lcccc}
Fit Range    &\multicolumn{2}{c}{$\leq425$ keV } &\multicolumn{2}{c}{$\leq1200$ keV}\\
\hline
Experiment\footnotemark   & Value & Error & Value & Error \\
\hline
Filippone        & 20.0 & 2.4 & 18.8 & 2.2\\
Hammache     & 19.4 & 1.2 & 18.8 & 1.0\\
Hass         &      &     & 19.7 & 1.0\\
Strieder     & 18.1 & 1.7 & 17.5 & 1.5\\
Baby             & 20.1 & 1.3 & 20.5 & 0.7\\
This work    & 21.4 & 0.6 & 21.5 & 0.6\\
\hline
${\leq}$425 keV best fit  & 20.6 & 0.5 &     &\\
${\leq}$1200 keV best fit &      &     &20.3 & 0.4\\
\end{tabular}
\end{ruledtabular}
\footnotetext[1]{Individual S$_{17}(20)$ values were computed from the original data before rounding.}
\end{table}

\begin{figure}
\includegraphics[width=0.5\textwidth]{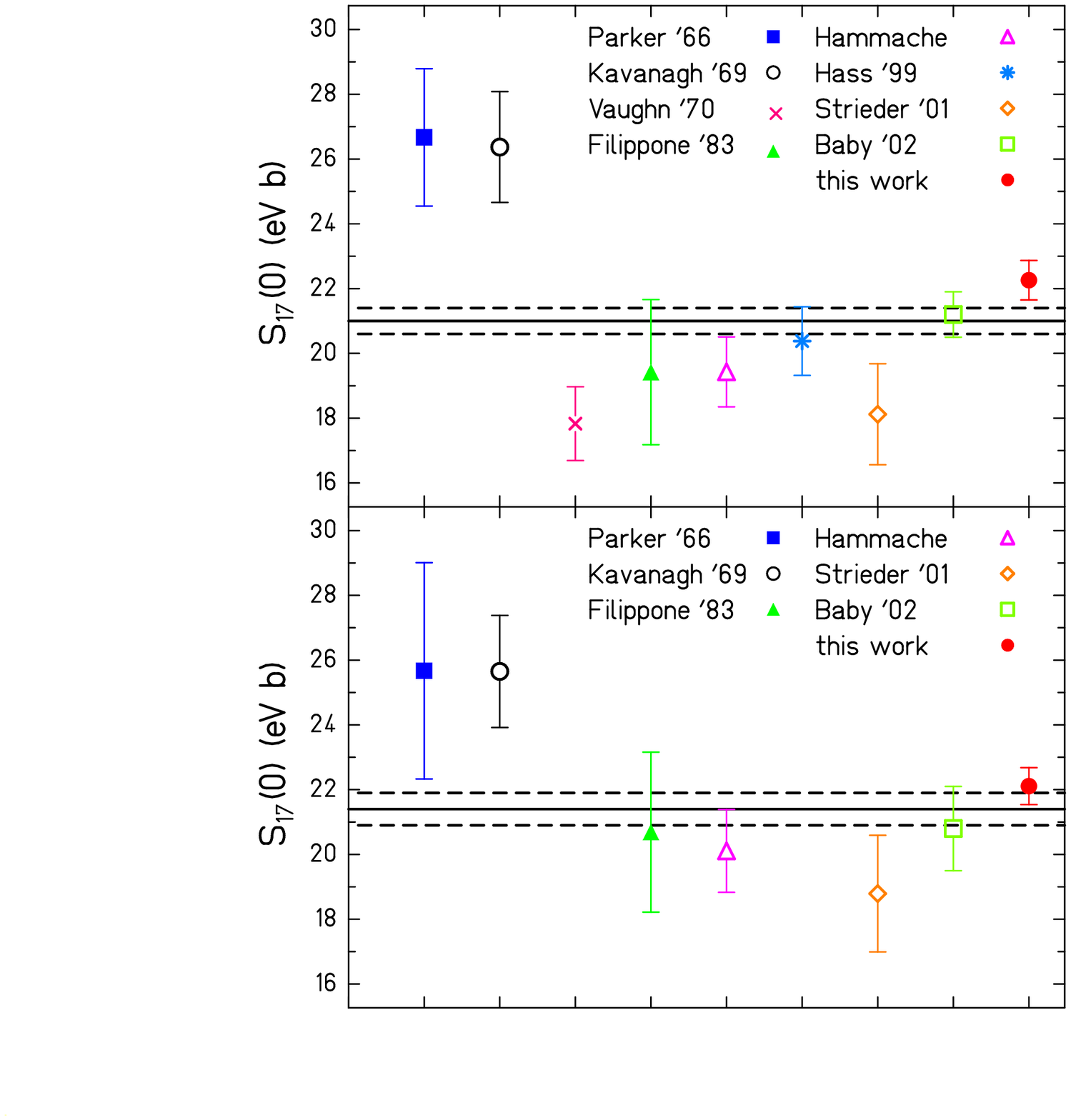}
\caption{$S_{17}(0)$ values determined from our DB fits to published data from direct experiments.  Bottom panel (top panel): fits to data with $\bar{E}_{\rm cm}$ $\leq$ 425 keV (1200 keV).  The horizontal solid and dashed lines indicate the mean values and uncertainties determined from fitting the data of Filippone and more recent experiments.}
\label{directS17comparison}
\end{figure}

We renormalized all published data that used the $^7$Li(d,p)$^8$Li normalization to  $\sigma$[$^7$Li(d,p)$^8$Li] = 152 $\pm$ 6 mb, the average of the results quoted in \cite{adelberger,weissman} for the cross section at the peak of the broad 780 keV resonance.   The results~\cite{hammache2} for both fitting ranges are shown in Fig.~\ref{directS17comparison} and Table~\ref{otherdirectS17values}.

Results from~\cite{parker,kavanagh,vaughn,filippone} may suffer additional error
from $^8$B and $^8$Li backscattering losses.  We estimated these losses for Filippone's experiment \cite{filippone} using TRIM and the target compositions described in~\cite{filippone2}.  We calculated both the loss of $^8$B from $^7$Be(p,$\gamma$)$^8$B and the loss of $^8$Li from the $^7$Li(d,p)$^8$Li reaction, the latter having been used by \cite{filippone} for one of the two absolute cross-section determinations.  We find smaller corrections by a factor of 2 compared to those published by Weissman ~\cite{weissman}.  Our calculated corrections for the weighted average of Filippone's normalizations, approximately  $-2$\% to $-4$\% depending on proton energy, are sufficiently small that we ignore them.  Hammache et al. \cite{hammache} applied calculated backscattering corrections to their data;  Strieder et al.\cite{strieder} used a low-Z backing and assumed negligible losses; and Baby et al. \cite{baby} used an implanted target for which losses should have been negligible.  We made no corrections to any of these published data for backscattering losses.

We combined all the direct S-factor data shown in Fig.~\ref{directS17comparison} except for the older pioneering experiments of Parker~\cite{parker}, Kavanagh~\cite{kavanagh} and Vaughn~\cite{vaughn}.
The results for $\bar{E}_{\rm cm} \leq$ 425 keV are,
S$_{17}$(0) = 21.4 $\pm$ 0.5 eV b, $\chi^2 / \nu$ = 1.2 ($\nu$ = 4), and for $\bar{E}_{\rm cm} \leq$ 1200 keV, S$_{17}$(0) = 21.1 $\pm$ 0.4 eV b, $\chi^2 / \nu$ = 2.1 ($\nu$ = 5).  These best fit values and uncertainties are listed also in Table~\ref{otherdirectS17values}.  The fit to the low-energy region has a good $\chi^2$, with $P(\chi^2,\nu)$ = 30\%.  The wide-range fit has $P(\chi^2,\nu)$ = 0.06, which is not unreasonable, though it suggests that some of the experimental uncertainties may be underestimated.  The good agreement between the S$_{17}(0)$ values determined from the 2 different fit ranges demonstrates that the energy dependence of the $^7$Be(p,$\gamma$)$^8$B cross section is well-understood and similar for all the direct experiments, and that the S$_{17}(0)$ value determined here is insensitive to the fit range.  We show in Table~\ref{otherdirectS17(20)values} the best-fit DB results for S$_{17}(20)$, analogous to Table~\ref{otherdirectS17values} for S$_{17}$(0).

As argued above, low-energy data provide the most reliable basis for determining S$_{17}$(0).  Hence our best value determined from direct experiments is
\begin{equation}
\mbox{S}_{17}(0) = 21.4 \pm 0.5 \mbox{(expt)} \pm 0.6 \mbox{(theor)}
\hspace{0.2cm} \mbox{eV b,}
 \label{bestdirectSfactor}
\end{equation}
where we have taken the theoretical extrapolation error from Sec.\ \ref{extrapolation}.

The energy dependence of the $^7$Be(p,$\gamma$)$^8$B  cross section is also well-determined by our measurements, since we have minimized all of the important uncertainties here.  Our data have small statistical uncertainties and small point-to-point scatter, both above and below the 1$^+$ resonance.  Our small area target - uniform beam flux technique allowed us to determine the number of beam-target interactions reliably for different beam energies and beam focussing conditions.  Our frequent {\em in situ} activity measurements determined the number of $^7$Be target atoms present during each (p,$\gamma$) measurement, avoiding problems due to losses from beam sputtering.  The bombarding-energy-dependent correction factor $F_{\alpha}(E_p)$ was minimal for our BE3 S detector data.  Our precision target profile measurements determined the effects of proton beam energy averaging in the target at each bombarding energy, and provided essential information on the stability of the target over time.

\begin{figure}
\includegraphics[width=0.5\textwidth]{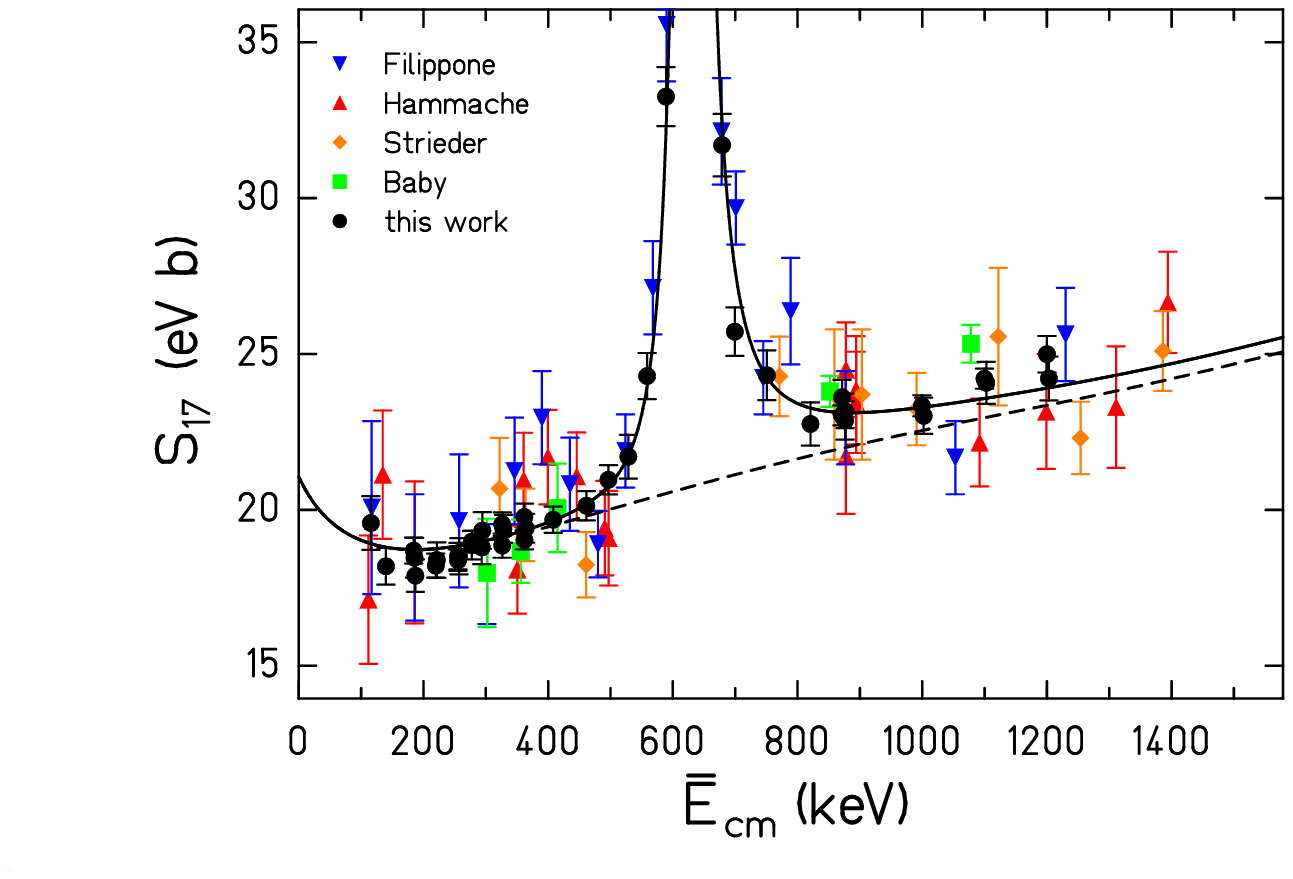}
\caption{S-factor data from direct experiments, all normalized to a common value of S$_{17}$(0) (the mean DB best-fit value of 21.1 eV b - see Table~\protect\ref{otherdirectS17values}).  The error bars shown are relative, and do not include scale-factor uncertainties.  Solid curve: DB plus a 1$^+$ resonance with parameters determined from fitting our BE1 data.  Dashed curve:  DB only.   Calculations and data are normalized in the energy range $\bar{E}_{\rm cm} \leq$ 1200 keV.}
\label{directenergydependence}
\end{figure}

We showed earlier (see Fig.\ref{BE1andBE3Sfactor}) that our present $^7$Be(p,$\gamma$)$^8$B results agree well with DB plus a fitted 1$^+$ resonance below 1200 keV or so.  In Fig.~\ref{directenergydependence} we compare the energy dependence of our results with the data from the other 4 modern direct experiments, all plotted with a common normalization corresponding to S$_{17}$(0) = 21.1 eV b based on our DB (plus a 1$^+$ resonance) fitted to data below 1200 keV (see Table~\ref{otherdirectS17values}).
From Fig.~\ref{directenergydependence} it can be seen that all direct experiments are in reasonable agreement on the energy dependence of the $^7$Be(p,$\gamma$)$^8$B  cross section
in the energy range $\bar{E}_{\rm cm} \leq$ 1200 keV.
This provides an important basis for comparing the direct results with indirect Coulomb-dissociation experiments (see below).

\subsection{Comparison with indirect experiments}
\label{indirectcompare}

\subsubsection{Coulomb dissociation}
Coulomb dissociation (CD) experiments, in which a secondary radioactive beam of $^8$B nuclei is dissociated into $^7$Be + p in the field of a heavy nucleus such as Au or Pb, have now been performed by several groups~\cite{motobayashi, kikuchi, iwasa, davids, schumann}.
In these experiments, one attempts to avoid nuclear contributions by measuring at very small $^8$B scattering angles.  Corrections for the branch to $^7$Be$^*$(429 keV) are required.  The relative weighting of E1, M1 and E2 multipolarities in the virtual-photon spectrum responsible for the Coulomb breakup is very different than the weighting in the direct photon-emission spectrum.  Hence, knowledge of the virtual-photon spectrum as well as the multipole decomposition of the direct $^7$Be(p,$\gamma$)$^8$B cross section are both needed to translate the measured breakup cross section into an equivalent direct cross section.  This is done with the assumption that the direct cross section, excluding the $1^+$ resonance, is pure E1.  In the virtual-photon spectrum, the E2 cross section, which is negligible in the direct process, is enhanced relative to E1 by several orders of magnitude  and  may not be negligible in the breakup cross section.  E2 contributions estimated from measured breakup momentum distributions range from small but significant~\cite{davids} to negligible~\cite{schumann}, and are not given reliably by theory.  The effect of the 630 keV $1^+$ M1 resonance is taken into account, while weaker M1 strength located at higher energies (see e.g. the 3$^+$ resonance discussion in Sec.~\ref{3resonance} above) is not treated explicitly.  Three-body Coulomb postdecay acceleration effects are a concern, especially at low relative $^7$Be + p energy and low bombarding energy.  A recent estimate~\cite{alt} suggests this effect may not be important in the work of refs.~\cite{iwasa, schumann}, while for the 81 MeV/nucleon, small-angle data of ref.~\cite{davids},  corrections for this effect have opposite signs for the points at  $\approx$ 200 and 400 keV relative $^7$Be + p energy.

\begin{figure}
\includegraphics[width=0.5\textwidth]{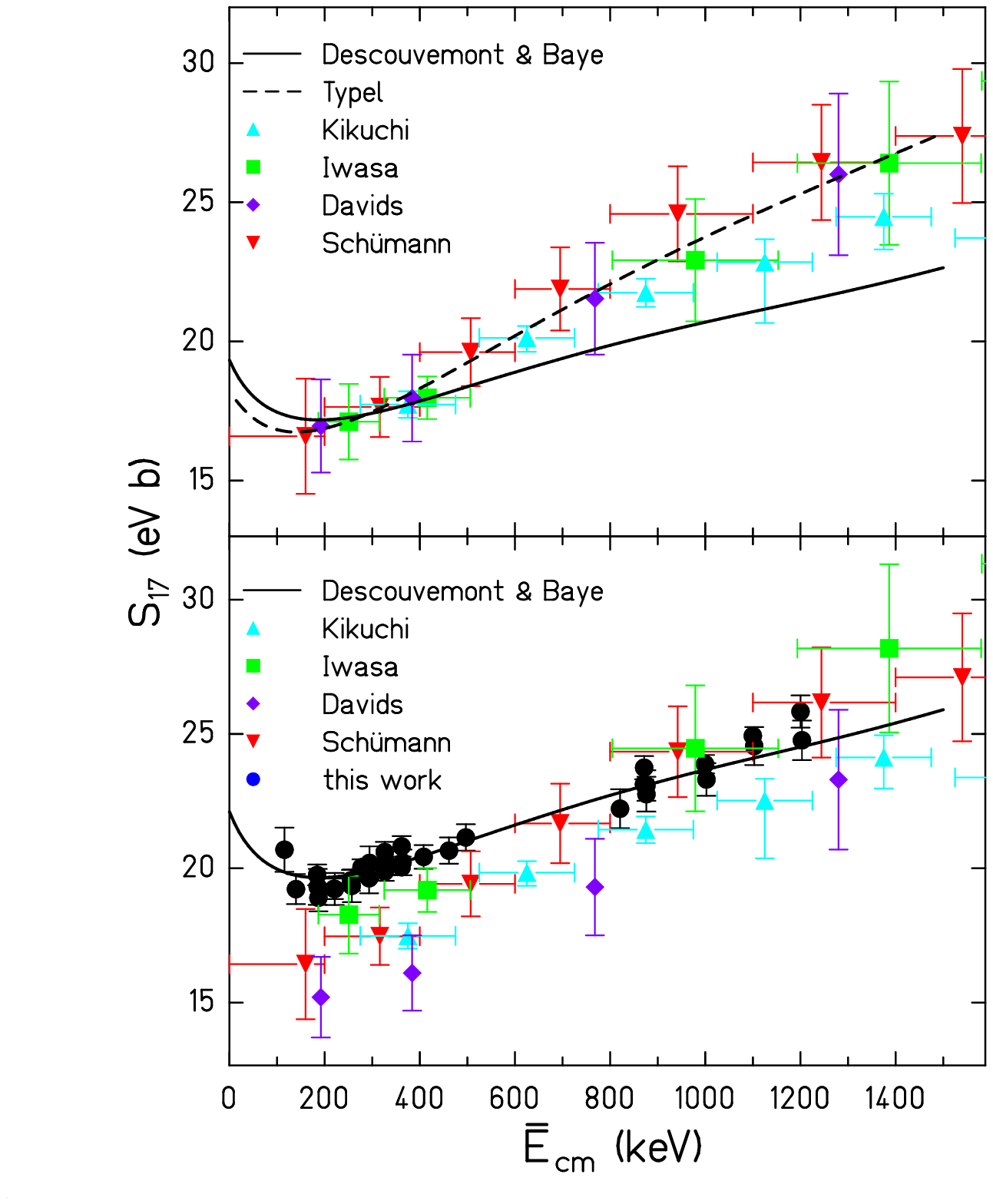}
\caption{E1 $^7$Be(p,$\gamma$)$^8$B S-factors inferred from Coulomb dissociation (CD) experiments.  Bottom panel: absolute CD S-factors, together with our direct results (with the 1$^+$ resonance subtracted) and the best-fit DB curve to our direct low-energy data.  Top panel: CD data plotted with a common normalization based on the mean value of 19.3 eV b for $S_{17}(0)$ determined by fitting each data set to the DB theory below 400 keV.  Solid curve: DB calculation; dashed curve: Typel calculation. The experimental error bars shown in all cases are relative, and do not include scale-factor uncertainties.}
\label{cdenergydependence}
\end{figure}

The energy dependence of the inferred E1 $S_{17}$-factors for the $^7$Be(p,$\gamma$)$^8$B reaction  are shown in the top panel of Fig.~\ref{cdenergydependence}, determined from the 4 most recent CD experiments. To display the energy dependence clearly,  we normalized each CD experiment to a common  value of  $S_{17}(0)$ determined by fitting each CD data set below 400 keV with the DB theory  (see Fig. 4a of \cite{schumann} for a similar plot with the original CD data, not renormalized).  The DB and Typel theory curves are also shown, again normalized to the CD data below 400 keV.   As can be seen, the energy dependence of the inferred $^7$Be(p,$\gamma$)$^8$B cross sections is similar in the works of Iwasa, Davids and Schumann, and is similar to the Typel theory, while the results of Kikuchi show a somewhat flatter energy dependence.

It should be noted that theory cannot predict reliably the energy dependence of the $^7$Be(p,$\gamma$)$^8$B cross section.  Hence the Typel calculation is useful here only as a guide to the eye that reproduces the energy dependence of the CD data.

We showed above that our direct measurements below 1200 keV have an energy dependence that is well-described by the DB theory
(plus a 1$^+$ resonance) except at the highest energies (1100 - 1200 keV) where DB falls a few percent low.   The energy dependence inferred from the CD experiments is {\em significantly steeper} than the DB theory and
does not agree with the direct results.
The bottom panel of Fig.~\ref{cdenergydependence} shows the CD data, not renormalized, along with our present direct results (with the 1$^+$ resonance subtracted) and the best-fit DB curve to the present data.

\begin{figure}
\includegraphics[width=0.5\textwidth]{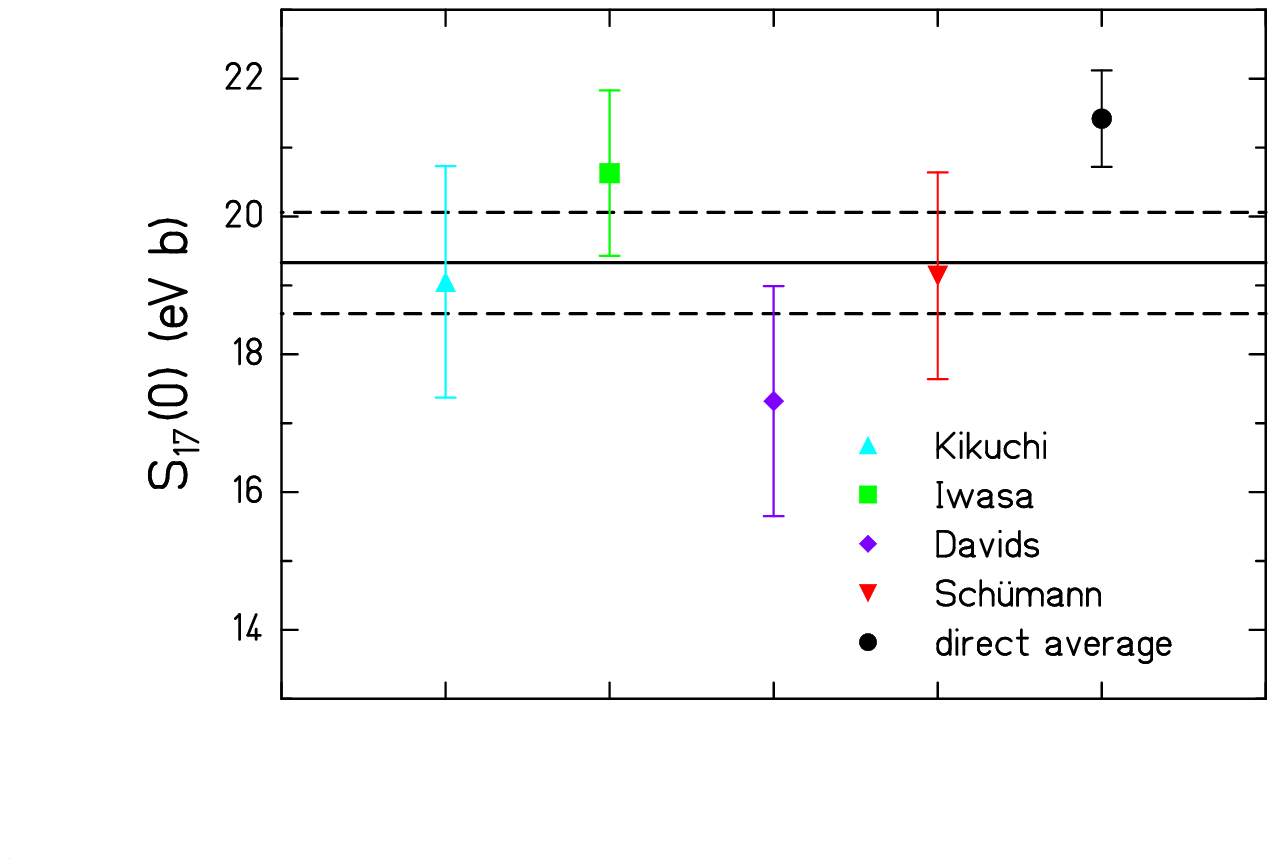}
\caption{CD $S_{17}(0)$ values from DB fits to inferred $^7$Be(p,$\gamma$)$^8$B cross sections below 400 keV.  The horizontal solid and dashed lines indicate the mean value  $S_{17}(0)$ = 19.3 $\pm$ 0.7 eV b.}
\label{cdS17values}
\end{figure}

Because of the different energy dependences observed in CD and direct experiments, it is difficult to know how to proceed with a more quantitative comparison.  If we ignore this problem, and focus on CD data below 400 keV, as we did with the direct experiments, then our DB fits yield the $S_{17}(0)$ values shown in Fig.~\ref{cdS17values}.  These values are mutually consistent, with a mean of 19.3$\pm$ 0.7 eV b.  A fit to this mean value together with the mean value deduced from direct experiments of 21.4 $\pm$ 0.5 eV b has probability P $\sim$ 0.02 that these results arise from the same parent distribution.  On the other hand, if we fit the CD data between 750 - 1400 (or 1000 - 1200) keV with the DB theory, the mean CD value is $S_{17}(0)$ $\approx$ 22 eV b, in very good agreement with the direct result (see Fig.~\ref{cdenergydependence}, bottom panel).  However, there seems to be no independent motivation for fitting only high-energy CD data.

\subsubsection{Heavy-ion transfer and breakup}

A Texas A\&M group has used measurements of peripheral heavy-ion transfer and breakup cross-sections to
deduce the asymptotic normalization coefficient for the $^7$Be + p component of the $^8$B ground state wavefunction.  This coefficient, together with a capture-model calculation  (and an assumed p$_{3/2}$/p$_{1/2}$ ratio in $^7$Be + p) can be used to infer S$_{17}(0)$.   The value S$_{17}(0)$ = 17.3 $\pm$ 1.8 eV b has recently been inferred from the weighted average of $^{10}$B($^7$Be, $^8$B)$^9$Be and $^{14}$N($^7$Be, $^8$B)$^{13}$C results at E($^7$Be) = 85 MeV~\cite{azhari}, and a variety of peripheral heavy ion breakup results at 28 to 285 MeV/u have been used to infer S$_{17}(0)$ = 17.4 $\pm$ 1.5 eV b~\cite{trache}.   However, a different analysis~\cite{brown} of the same breakup reaction measured with a C target leads to a substantially larger S$_{17}(0)$ value of 21.2 $\pm$ 1.3 eV b in good agreement with the direct mean value of 21.4 eV b.  A determination of the asymptotic normalization coefficients for the p$_{1/2}$ and p$_{3/2}$ components of $^8$Li $\rightarrow$ $^7$Li + n together with the assumption of mirror symmetry leads to S$_{17}(0)$ = 17.6 $\pm$ 1.7 eV b for $^7$Be + p~\cite{trache2}.  These S-factor determinations thus tend to be even smaller than those deduced from CD experiments.

S(E) values inferred from  $^{16}$O($^3$He,d)$^{17}$F cross section measurements~\cite{gagliardi} have been compared to direct $^{16}$O(p, $\gamma$)$^{17}$F  cross section measurements~\cite{chow,morlock} for capture to the ground state and to the first excited state.  For the first-excited state transition, the (p, $\gamma$) results of ref.~\cite{chow} and the transfer-reaction results agree to approximately (6 $\pm$ 11)\%, where the uncertainty is determined by the $\pm$ 10\%  systematic uncertainty in the transfer reaction and $\pm$ 5\% uncertainty in the absolute (p, $\gamma$) cross section.  For the ground-state transition, the central values from (p, $\gamma$)~\cite{morlock}and from ($^3$He,d) agree within 10\% or so but it is difficult to quantify the significance of the comparison\cite{gagliardi} since the absolute (p,$\gamma$)cross section uncertainty was not specified.

\section{Recommended values for S$_{17}$(0) and S$_{17}$(20)}
\label{recommendvalue}
In comparing results from experiments that employ very different techniques, it is necessary to group the results according to technique, as we have done above, to see if the results are technique-dependent.  The evidence presented here points strongly toward such a technique dependence.

One could obtain a ``best" or recommended value for  S$_{17}$(0) by combining the mean results from different techniques, expanding the error on each mean by a common factor such that the combined fit has probability P = 50\%; i.e. $\chi^2/\nu$ = 0.46 for $\nu$ = 1.  With this procedure, the combination of direct plus low-energy CD results yields S$_{17}$(0) = 20.7 $\pm$ 1.5 (expt) $\pm$ 0.6 (theor) eV b.  Including the heavy ion reaction results as a separate, third group would increase the overall uncertainty and further lower the central value.
This procedure has the disadvantage that it treats the results from different techniques on an equal footing.
We have shown that the $^7$Be(p,$\gamma$)$^8$B energy dependence inferred from CD experiments disagrees with direct results.  In our opinion, this difference in energy dependence must be understood before results from direct and indirect experiments can be combined.  In addition, considerable theoretical modeling is necessary to infer the $^7$Be(p,$\gamma$)$^8$B cross section from CD experiments, or S$_{17}$(0) from heavy-ion transfer and breakup experiments, and it is difficult to understand all the uncertainties associated with this modeling.

It is important to note that neither indirect technique has been tested  by comparison to a known direct result with sufficient precision to demonstrate that systematic uncertainties are understood at the level of $\pm$ 3 - 5\%.
We conclude that, at present, the indirect experiments are not sufficiently understood to be included in the determination of a recommended value.

A new direct $^7$Be(p,$\gamma$)$^8$B cross section measurement using a $^7$Be beam would be useful as an independent determination of the absolute cross section with systematic uncertainties different from those of $^7$Be target experiments.  However, to make a significant contribution, such a measurement would have to have a total experimental uncertainty of 5\% or better.

We base our recommendation for S$_{17}$(0) on the mean of direct experiments as given in Eq.\ \ref{bestdirectSfactor} above, which we repeat here:
\begin{equation}
\mbox{S}_{17}(0) = 21.4 \pm 0.5 \mbox{(expt)} \pm 0.6 \mbox{(theor)}
\hspace{0.2cm} \mbox{eV b,}
 \label{rec1}
\end{equation}
where the quoted errors are 1$\sigma$.
Although our discussion focusses on S$_{17}$(0), it seems clear that theoretical extrapolation uncertainties should ultimately be reduced by using instead the S-factor at  20 keV \cite{jennings}, close to the Gamow peak. Our recommendation for this quantity and its 1$\sigma$ uncertainty is
\begin{equation}
\mbox{S}_{17}(20) = 20.6 \pm 0.5 \mbox{(expt)} \pm 0.6 \mbox{(theor)}
\hspace{0.2cm} \mbox{eV b.}
\label{rec2}
\end{equation}

\section{Summary}
\label{summary}

We made new $^7$Be(p,$\gamma$)$^8$B cross section measurements that extend our earlier results to lower energy and have improved systematic errors.   Based on our new data with $\bar{E}_{\rm cm}$ = 116 to 362 keV and the cluster-model theory of Descouvemont and Baye, we determine S$_{17}$(0) = 22.1 $\pm$ 0.6 (expt) $\pm$ 0.6 (theor) eV b, where the theoretical (extrapolation) error is given by the rms deviation of S$_{17}$(0) values from 12 different theories fitted to the same data.  Our new result is in excellent agreement with our previously published value of 22.3 $\pm$ 0.7(expt) eV b~\cite{junghans}, and supercedes it.

We have fitted all published direct  $^7$Be(p,$\gamma$)$^8$B measurements at $\bar{E}_{\rm cm} \leq$ 425 keV and $\bar{E}_{\rm cm} \leq$ 1200 keV with the DB theory.  The modern experiments (Filippone, Hammache, Strieder, Baby and this work) give very consistent results.
For $\bar{E}_{\rm cm} \leq$ 425 keV , the combined fit yields S$_{17}$(0) = 21.4 $\pm$ 0.5 (expt) $\pm$ 0.6 (theor) eV b, with $\chi^2/\nu$ = 1.2 (P = 30\%).
At present, the uncertainties in S$_{17}$(0) from the experimental cross sections and from the theoretical extrapolation are nearly the same. This points to the importance of further theoretical work to reduce the extrapolation uncertainty.

We have also examined the 4 recent Coulomb dissociation experiments.  These experiments infer a steeper energy dependence for the $^7$Be(p,$\gamma$)$^8$B cross section below 1200 - 1500 keV than is observed in the direct experiments.  Since the energy dependence of the present  direct measurements is unambiguous (at the level of a few percent), this indicates a systematic error in the interpretation of the CD experiments.  Fitting the CD data below 400 keV with the DB theory leads to a mean value of  S$_{17}$(0) = 19.3 $\pm$ 0.7 (expt) $\pm$ 0.6 (theor) eV b, which is only compatible with the mean of direct measurements at the level of 2\%.  Fitting only the high-energy region of these data yields S$_{17}$(0) $\approx$ 22  eV b, in excellent agreement with the direct result; however, it is not clear that such a restricted fit is well-motivated.  Peripheral heavy-ion transfer and breakup experiments lead to S$_{17}$(0) $\approx$ 17.4 eV b.

Our recommended values for S$_{17}(0)$ and S$_{17}$(20), based on the totality of existing data, are given in Eqs.~\ref{rec1} and \ref{rec2}.

\section{Impact of new S$_{17}$(0) on the solar model and on neutrino physics}
\label{neutrinoimplications}

Our recommended S$_{17}$(0) has a combined experimental plus theoretical uncertainty of $\pm$ 0.8 eV b, or $\pm$ 4\%.  This represents a considerable improvement over the 1998 recommendation by Adelberger et al.~\cite{adelberger} of S$_{17}$(0) = 19 $^{+ 4}\!\!\!\!\!\!_{- 2}$ eV b (1$\sigma$).  It also represents a considerable improvement compared to S$_{17}$(0) = 19 $^{+ 8/3}\!\!\!\!\!\!\!\!\!\!\!_{- 4/3}$ eV b assumed in BP00~\cite{adelberger2}.   If incorporated into the BP00 solar model calculation~\cite{bahcall}, the uncertainty in S$_{17}$(0) would no longer make an important contribution to the overall uncertainty in the calculated solar neutrino production rate from $^8$B decay, and the overall uncertainty would be reduced from $\pm$ 17\% to $\pm$ 14\%.

Recent combined analyses of solar-neutrino plus initial KamLAND results have limited the allowed oscillation parameters to the LMA (large mixing angle) region~\cite{holanda,bahcall3}.  These analyses, which are independent of the solar model, determine  $f_{\rm B,total}$ = 1.00 $\pm$ 0.06~\cite{bahcall3}, or 1.05 -1.08~\cite{holanda} (no uncertainty quoted), depending on method of analysis, where $f_{\rm B,total}$ is the total $^8$B neutrino flux (at the surface of the earth) in units of the BP00 flux.  Since the SSM $^8$B neutrino flux depends linearly on S$_{17}$(0), incorporating our recommended  S$_{17}$(0) value and uncertainty into the SSM leads to a new SSM $^8$B neutrino flux  $f_{\rm B,total}$ = 1.13 $\pm$ 0.16, in units of the BP00 flux.  Thus the correctness of the SSM has been dramatically confirmed.

Within 3$\sigma$, these solar-model-independent analyses also restrict the allowed LMA region to a primary minimum at $\Delta m^2$ $\approx$ 7 x 10$^{-5}$ eV$^2$ and a secondary minimum at  $\Delta m^2$ $\approx$ 1.5 x 10$^{-4}$ eV$^2$.  In the analysis of ref.~\cite{bahcall3} these minima lie at $f_{\rm B,total}$ = 1.00 and 0.88, respectively (see Table 3 and Fig. 4 of \cite{bahcall3}).  Thus the disfavored (secondary) minimum with $f_{\rm B,total}$ = 0.88 is somewhat disfavored additionally by the SSM rate calculated with our recommended S$_{17}$(0).

Because the combined atmospheric neutrino, solar neutrino-plus-KamLAND, and LSND results require 3 distinct regions of allowed $\Delta m^2$, a fourth (sterile) light neutrino is required if all 3 data sets are correct.   However, even if the LSND results turn out to be spurious, sterile neutrinos may exist and play important roles in nature
(see e.g. \cite{holanda2}).
Existing limits on sterile neutrinos are derived from
solar-neutrino plus Kamland-antineutrino data~\cite{bahcall3} assuming CPT invariance. But the CPT properties of sterile neutrinos, unlike those of other particles, have not been tested.  A rigorous test for sterile neutrinos requires knowing the production and detection rates for at least two {\em neutrino} sources with substantially different energy spectra. These could be the pp neutrinos (whose production rate is determined by the solar luminosity) and the solar $^8$B neutrinos
whose production rate is affected by the results of this work.
The new SSM $^8$B flux based on our recommended S$_{17}$(0) would be useful in an analysis that tests for sterile neutrinos without assuming CPT invariance.

We thank the staff of CENPA, particularly G. C. Harper and J. F. Amsbaugh, the staff of TRIUMF, and N. Bateman, R. Hoffenberg, J. Martin, L. Melling, N. Matsuda,  R. O'Neill, P. Peplowski and C. Rivet
for their help.
The U.S. D.O.E., Grant $\#$DE-FG03-97ER41020, and the N.S.E.R.C. of Canada
provided financial support.


\begin{thebibliography}{}
\bibitem{be22label} This is the same target as the one labelled BE22 in ref.~\protect\cite{zyuzin2}.
\bibitem{french} The $^{137}$Cs source was fabricated at the French CERCA laboratory, and its activity was calibrated at LNHB, CEA, Saclay, France.
\bibitem{hammache2} For the data of \protect\cite{hammache} we computed an average systematic error from the errors quoted on the '95 and '96 data, and treated this error as common mode.  We then combined the '95+'96 result and the '98 result for S$_{17}(0)$, treating these 2 results as independent.
\bibitem{adelberger2}
This uncertainty is 1/3 of the ``3$\sigma$" uncertainty quoted in Eqn. 29 of \protect\cite{adelberger}.
\end{thebibliography}
\end{document}